\documentclass[prl,twocolumn,amsmath,amssymb,superscriptaddress,showpacs,numbers,openany]{revtex4-2}

\usepackage[utf8]{inputenc}
\usepackage{color}
\usepackage{latexsym}
\usepackage{float,ulem}
\usepackage{dsfont}
\usepackage[dvipsnames]{xcolor}
\usepackage[english]{babel}
\usepackage{lipsum}

\usepackage{tabularx} 
\usepackage{booktabs}

\usepackage{bm, mathtools}
\usepackage{graphicx}
\usepackage{dcolumn}
\usepackage{times}
\usepackage[english]{babel}
\usepackage[T1]{fontenc}
\usepackage{gensymb}

\usepackage{cellspace}%
\usepackage{makecell}
\setcellgapes{3pt}
\newcommand{\fig}[1]{Fig.~\ref{fig:#1}}

\newcommand{\dc}{\degree \text{C}}
\newcommand{\tcr}[1]{\textcolor{black}{#1}}

\definecolor{linkcolor}{rgb}{0,0,0.6} 
\usepackage[pdftex,colorlinks=true, pdfstartview=FitV, linkcolor= linkcolor, citecolor= linkcolor, urlcolor= linkcolor, hyperindex=true,hyperfigures=true]{hyperref}

\newcommand{\be}{\begin{equation}}
\newcommand{\ee}{\end{equation}}

\begin{document}

\setcounter{secnumdepth}{1}

\title{Phase behavior of Cacio e Pepe sauce}

\author{G. Bartolucci}
\thanks{All authors contributed equally to this work}
\affiliation{Department of Physics, Universitat de Barcelona, Barcelona, Spain}
\author{D. M. Busiello}
\thanks{All authors contributed equally to this work}
\affiliation{Max Planck Institute for the Physics of Complex Systems, Dresden, Germany}
\affiliation{Department of Physics and Astronomy ``G. Galilei'', University of Padova, Padova, Italy}
\author{M. Ciarchi}
\thanks{All authors contributed equally to this work}
\affiliation{Max Planck Institute for the Physics of Complex Systems, Dresden, Germany}
\author{A. Corticelli}
\thanks{All authors contributed equally to this work}
\affiliation{Max Planck Institute for the Physics of Complex Systems, Dresden, Germany}
\author{I. Di Terlizzi}
\thanks{All authors contributed equally to this work}
\affiliation{Max Planck Institute for the Physics of Complex Systems, Dresden, Germany}
\author{F. Olmeda}
\thanks{All authors contributed equally to this work}
\affiliation{Institute of Science and Technology Austria, Am Campus 1, 3400 Klosterneuberg, Austria }
\author{D. Revignas}
\thanks{All authors contributed equally to this work}
\affiliation{Max Planck Institute for the Physics of Complex Systems, Dresden, Germany}
\affiliation{Department of Physics and Astronomy ``G. Galilei'', University of Padova, Padova, Italy}
\author{V. M. Schimmenti}
\thanks{All authors contributed equally to this work}
\affiliation{Max Planck Institute for the Physics of Complex Systems, Dresden, Germany}

\begin{abstract}
\noindent 
``Pasta alla Cacio e pepe'' is a traditional Italian dish made with pasta, pecorino cheese, and pepper. Despite its simple ingredient list, achieving the perfect texture and creaminess of the sauce can be challenging. In this study, we systematically explore the phase behavior of Cacio e pepe sauce, focusing on its stability at increasing temperatures for various proportions of cheese, water, and starch. We identify starch concentration as the key factor influencing sauce stability, with direct implications for practical cooking. Specifically, we delineate a regime where starch concentrations below 1\% (relative to cheese mass) lead to the formation of system-wide clumps, a condition determining what we term the ``Mozzarella Phase'' and corresponding to an unpleasant and separated sauce. Additionally, we examine the impact of cheese concentration relative to water at a fixed starch level, observing a lower critical solution temperature that we theoretically rationalized by means of a minimal effective free-energy model. \tcr{We further analyze the effect of a less traditional stabilizer, trisodium citrate, and observe a sharp transition from the Mozzarella Phase to a completely smooth and stable sauce, in contrast to starch-stabilized mixtures, where the transition is more gradual.} Finally, we present a scientifically optimized recipe based on our findings, enabling a consistently flawless execution of this classic dish.

\end{abstract}

\maketitle

\section{Introduction}

\begin{figure*}[!ht] 
\includegraphics[width=1\linewidth]{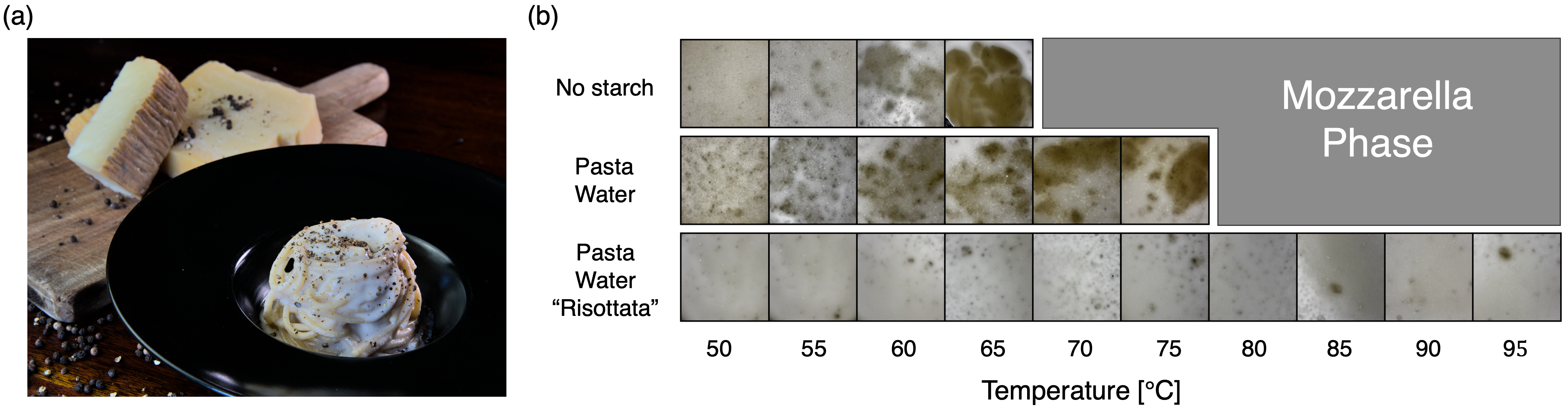}
\caption{\textbf{Cacio e pepe pasta sauce consists of pecorino cheese, pepper and starch-enriched water} (a) \tcr{Noodles with an emulsion of pecorino cheese and starch-enriched water, garnished with freshly ground black pepper.}
(b) Snapshots of the mixture that constitutes the base for the pasta sauce, i.e. cheese and water with different amounts of starch, at different temperatures. In particular, we compare the effect of: water 
alone; pasta water that retains some starch (obtained by cooking $100$ g of pasta in $1$ liter of water); and pasta water ``risottata'', i.e. pasta water heated in a pan to let the water evaporate (until reducing its total weight by three times) and starch gets concentrated. As the starch concentration increases, cheese aggregates decrease in size and occur at higher temperatures. The region here named ``Mozzarella Phase'' is characterized by huge mozzarella-like clumps of cheese suspended in water, resulting from extreme protein aggregation on heating.}
\label{fig:pasta}
\end{figure*} 

On several occasions, pasta has been a source of inspiration for physicists~\cite{ballPastaPhysics2010}.
The observation that spaghetti always break up into three or more fragments, but never in two halves, puzzled even Richard Feynman himself, and the explanation of this intriguing phenomenon earned Audoly and Neukirch the Ig Nobel Prize~\cite{audolyFragmentationRodsCascading2005}. 
Pasta packaging offers a natural framework to study separation upon shaking~\cite{caulkinGeometricAspectsParticle2010} and inspired the design of ``morphing flat pasta''~\cite{taoMorphingPasta2021}. Furthermore, the deformation and swelling behavior of various pasta varieties upon cooking has been experimentally and theoretically investigated~\cite{hwangSwellingSofteningElastocapillary2022, goldbergMechanicsbasedModelCookinginduced2020, gonzalezTexturalStructuralChanges2000}. 
\tcr{Lastly, analogies with pasta shapes proved useful in different physics fields, from polymer rings to neutron stars ~\cite{michielettoTasteAnelloni2014, caplanColloquiumAstromaterialScience2017}.}

\tcr{Pasta water also exhibits fascinating physical properties, primarily due to its starch content. When heated, starch-water solutions undergo a gelatinization transition~\cite{jenkinsGelatinisationStarchCombined1998, takoPrinciplesStarchGelatinization2014}, altering their viscosity and structural properties. Additionally, mixing water with 1.5–2 parts of corn starch is a well-known method for creating a non-Newtonian fluid, often referred to as oobleck~\cite{krishnaExperimentalEvaluationImpact2021}. Starch-enriched water also plays a crucial role in stabilizing emulsions, as seen in the classic dish ``spaghetti aglio e olio'', where it helps form a smooth, creamy sauce by preventing the separation of oil into suspended droplets.}

\tcr{In the culinary realm, many other examples highlight the importance of emulsions containing droplets formed via phase separation~\cite{mathijssenCulinaryFluidMechanics2023}. Phase separation often plays a crucial role in determining food texture~\cite{tolstoguzovTexturisingPhaseSeparation2006}, as seen in mayonnaise, salad dressings, and various sauces~\cite{fribergFoodEmulsions2003}. In such emulsions, the homogeneous state achieved by blending or shaking is metastable and naturally evolves toward a thermodynamically stable state characterized by the formation of oily droplets, often resulting in an undesirable consistency. To maintain a uniform texture, emulsions must be stabilized to slow down this ripening process~\cite{dickinsonFoodEmulsionsFoams2010}. 
The need to delay phase separation extends beyond emulsions to other food products, such as chocolate~\cite{koizumiControlPhaseSeparation2022} and ice cream~\cite{aichingerPhaseSeparationFood2017}, where unwanted phase separation leads to textural degradation and reduced shelf life. Droplet physics plays an important role in beverages as well. A famous example is the ouzo-effect, occurring in alcohols such as ouzo, pastis, and limoncello~\cite{vratsanosOuzoEffectExamined2023, grilloSmallangleNeutronScattering2003, chiappisiLookingLimoncelloStructure2018}.}

\tcr{In biochemistry, phase separation has recently received a new wave of enthusiasm caused by the discovery of membrane-less compartments in the cell cytoplasm~\cite{brangwynneGermlineGranulesAre2009}. These compartments have been successfully described as droplets formed via phase separation~\cite{hymanLiquidLiquidPhaseSeparation2014} and are associated with various essential cell functions~\cite{bananiBiomolecularCondensatesOrganizers2017, albertiConsiderationsChallengesStudying2019}. Furthermore, alterations of membrane-less compartments' physiological properties can trigger the formation of pathological protein aggregates~\cite{zbindenPhaseSeparationNeurodegenerative2020}. Meanwhile, the fascinating idea that phase separation could have played a role in the primordial soup, originally proposed by Oparin and Haldane~\cite{oparinOriginLife1952, haldaneOriginLife1929}, got reignited, thereby leading to new studies on mixtures composed of short oligomers~\cite{moraschHeatedGasBubbles2019,bartolucciSequenceSelfselectionCyclic2023}. }

Building upon the recent theoretical and experimental advances in the thermodynamics of mixtures, in this paper we investigate the phase behavior of Cacio e pepe pasta sauce. Cacio e pepe (literally ``cheese and pepper'') is a traditional recipe from Lazio, a region in central Italy. It consists of tonnarelli noodles served in a cream of pecorino cheese, pepper, and starch-enriched water, see \fig{pasta}(a).
\tcr{A precursor to this dish can be traced back to the fifteenth century, as documented in \cite{LiberCoquina}. The modern version, however, is commonly linked to the long journeys of shepherds, who had to fill their saddlebags with highly caloric and easily transportable ingredients to sustain them during their travels.}
Pecorino cheese was ideal due to its extraordinary shelf life, black pepper was used to stimulate heat receptors, and homemade \tcr{noodles} provided the carbohydrate intake. 

Despite the short list of ingredients, preparing this dish requires extra care. \tcr{This is because cheese and water mixtures are notoriously prone to separation when heated, as seen for example in Swiss cheese fondue \cite{bertschRheologySwissCheese2019}. In the context of Cacio e pepe, this challenge has already been explored by Dario Bressanini~\cite{bressaniniRicetteScientificheCacio}.}
The most delicate step is the mixing of starch-enriched water with grated cheese. One starts cooking pasta noodles in boiling water, letting them release starch, then extracts part of the water and starch solution. An essential procedure is to wait some time before mixing water and cheese, to let the water cool down. This is because, at high temperatures, cheese proteins can either form clumps upon denaturation or simply aggregate \cite{WheyMicelles1,WheyMicelles2}, therefore ruining the sauce, see \fig{pasta}(b). But temperature is not the only physical parameter that must be carefully controlled: protein aggregation and denaturation are concentration-dependent processes, thus, mixing the right amount of cheese, water, and starch is essential to avoid protein aggregates. In the absence of starch, for example, cheese in hot water forms huge clumps at a temperature around $65 \dc$, \tcr{leading to what we name the ``Mozzarella Phase''}, see the first row of \fig{pasta}(b). If cheese is mixed with pasta water, which contains small amounts of starch, the emergence of clumps is reduced, and large protein aggregates are found at higher temperatures, see \fig{pasta}(b), second row. Pasta water can be ``risottata'', i.e. collected and heated in a pan, so that some water evaporates and the starch is concentrated. If cheese is mixed with pasta water ``risottata'', the presence of clumps is almost negligible, see the last row in \fig{pasta}(b). 

To overcome clump formation and achieve the perfect Cacio e pepe, in this work, we characterize the phase behavior of the solution containing water, starch, and cheese. We achieve this using common kitchen tools, ensuring that our results are easily reproducible not only by scientists in the lab but also by culinary enthusiasts. The paper is organized as follows: in Sec. \ref{sec:vary_starch}, we introduce the experimental setup and discuss the role of starch in the mixture phase behavior. In Sec. \ref{sec:vary_water}, we fix the starch percentage and vary the cheese amount. Sec. \ref{sec:model} discusses a minimal theoretical model that recapitulates our experimental finding. \tcr{In Sec. \ref{sec:citrate}, we also investigate the role of a less traditional stabilizer, trisodium citrate, in preventing clump formation, analyzing its impact on sauce stability and comparing its effects to those of starch.} Finally, in Sec. \ref{sec:recipe}, we propose a recipe for the perfect Cacio e pepe.  

\section{Increasing starch mitigates the formation of protein aggregates}
\label{sec:vary_starch}

\begin{figure*}[!ht] 
\includegraphics[width=1\linewidth]{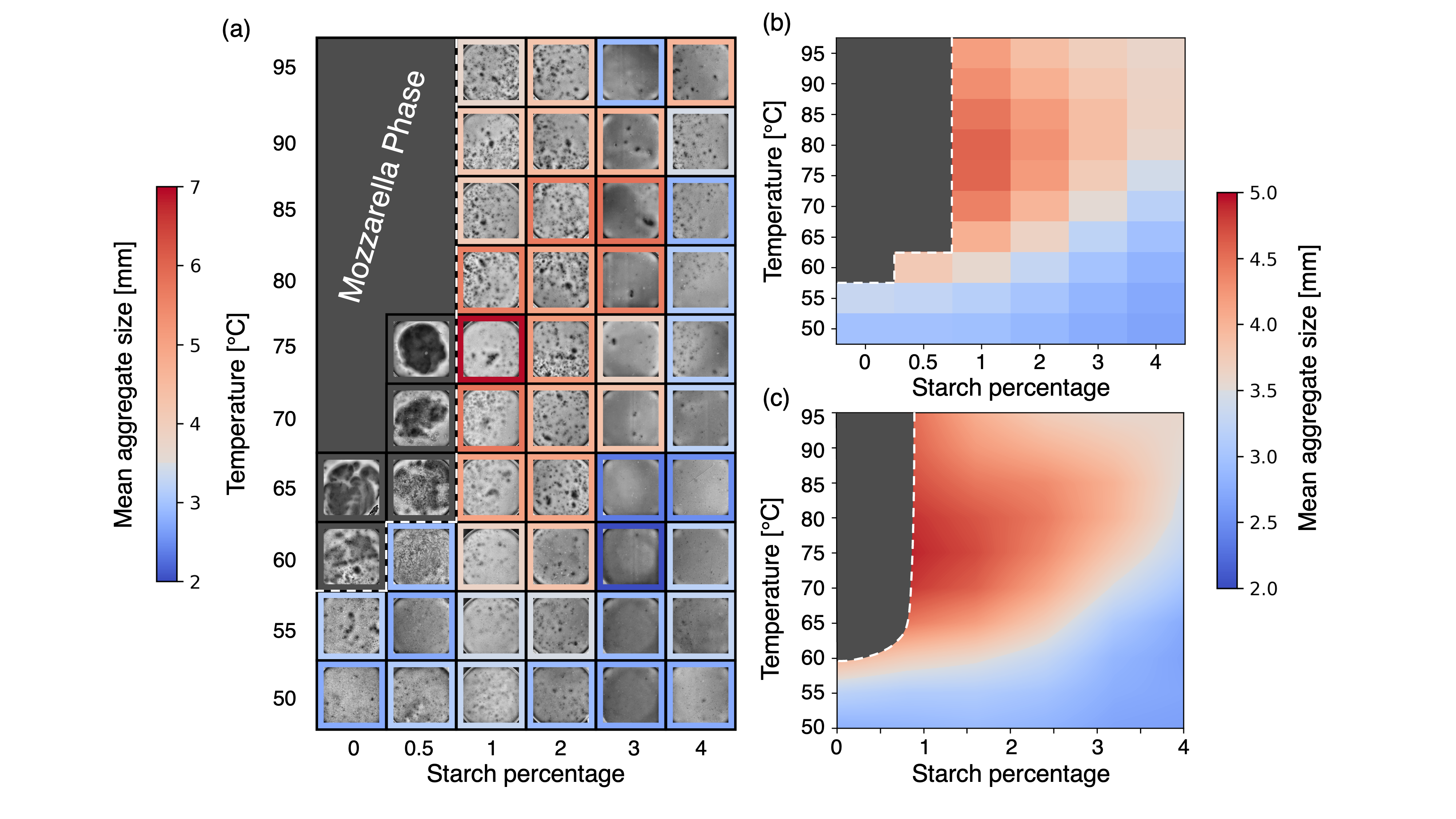}
\caption{\textbf{Starch mitigates protein clumps.} (a) Phase diagram of the sauce state as a function of the starch percentage with respect to the water and temperature in Celsius. Each box contains a snapshot of the sauce mixture taken during the experiment, and its contour reflects the mean size of the corresponding sample via the color map shown on the left. The Mozzarella Phase indicates a region of the phase diagram where the cheese in the sample forms a clump of a size comparable to the one of the system. (b) The same phase diagram from (a) after applying Gaussian smoothing to better visualize phase behavior. The color represents the mean aggregate size in the sample. (c) Kernel regression smoothing of the phase diagram in a) to obtain a continuous map. The color map on the right refers to panels b) and c).}
\label{fig:vary_starch}
\end{figure*} 

\begin{figure*}[!t] 
\includegraphics[width=1\linewidth]{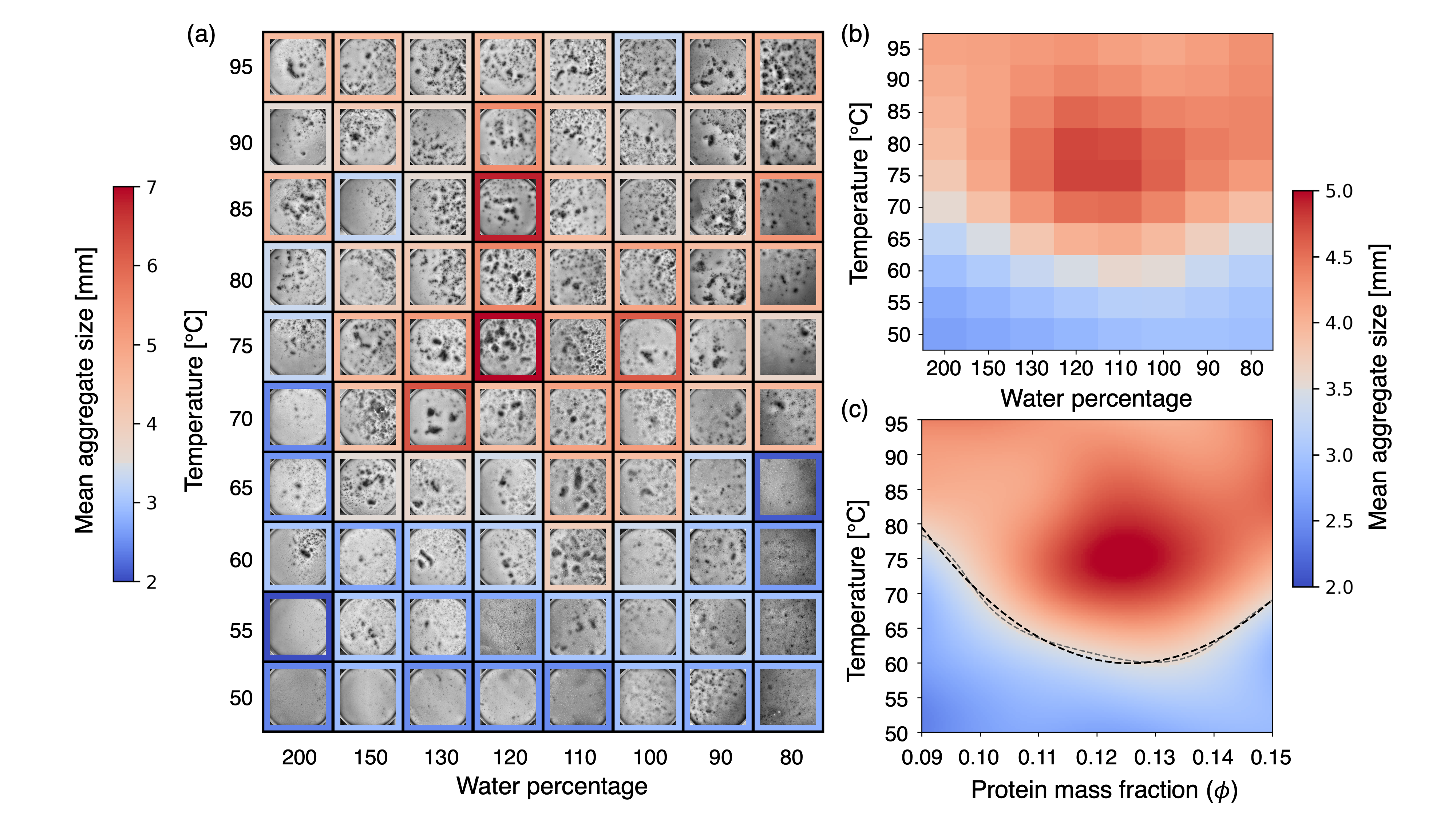}
\caption{\textbf{Protein amount tunes aggregate formation.} (a) Phase diagram of the sauce made by combining $50$ g of cheese with a varying amount of starch-enriched water (here expressed as a mass percentage) as a function of the mixture temperature. The starch percentage is fixed to $1$ \%. Each box is colored with respect to the mean size of cheese clumps. (b) Same phase diagram as in (a) after Gaussian smoothing, where each point is colored according to the map of the mean aggregate size (color map on the right). (c) Phase diagram expressed as a function of protein mass fraction $\phi$, which is the component that leads to aggregation. Kernel regression smoothing has been applied to obtain a continuous diagram. The dashed gray line indicates the isoline of median aggregate size, while the dashed black line represents the parabolic fit separating homogeneous from clumped domains.}
    \label{fig:vary_water}
\end{figure*}

As discussed above, one of the most problematic aspects of making Cacio e pepe is the appearance of large aggregates of cheese
(Mozzarella Phase). 
\tcr{The presence of starch is the main difference between pasta water and normal water, suggesting that starch concentration is an important factor in determining the formation of cheese clumps.} 

\tcr{In order to quantitatively investigate the role of starch}, we conducted experiments at a fixed cheese-to-water ratio while varying the starch concentration in the water and the temperature of the system. 
To ensure quality and reproducibility and to comply with the traditional recipe, we used a Pecorino Romano \tcr{DOP, whose nutritional values and certification details are discussed in the supplementary material, Sec. \ref{Ingredients} and Table \ref{tab:nutritional}.}

We prepared the starch-enriched water, dissolving dry corn starch into water at ambient temperature, targeting the weight percentages shown on the bottom axis of \fig{vary_starch}(a). The mixture was heated on a stovetop to gelatinization, a transition marked by a noticeable increase in viscosity and opacity. Afterward, we allowed the starch-enriched water to cool to room temperature to prevent excessive heating of the cheese-water mixture during blending. Cheese and starch-enriched water were combined in equal weights and thoroughly blended with a mixer. The resulting mixture was placed in a controlled heat bath, namely a pot of water maintained at a constant temperature using a \textit{sous vide} cooker device. 
The mixture's temperature was further monitored with an external thermometer positioned inside it, and samples were collected upon reaching thermal equilibrium, as indicated by a stable mixture temperature. 
Each sample was spooned onto a Petri dish and photographed. In the supplementary material, Sec. \ref{sec:S1}, we report on the experimental apparatus built to heat the mixture and collect images, along with a detailed explanation of the measurement protocol. 
Each mixture 
was imaged systematically increasing the heat bath temperature by $5$ \dc ~at each sampling and the evaporated water was replenished at each temperature step.
Experiments were performed at different starch percentages to generate the images forming the phase diagram in \fig{vary_starch}(a). \tcr{At low temperatures, we did not observe protein aggregates independently of the starch concentration. However, by increasing the temperature, clumps start being detected as dark areas in the pictures in \fig{vary_starch}(a), thanks to a light source below the Petri dish (see Figs.~\ref{fig:schema_apparato} and~\ref{fig:foto_apparato} in supplementary material).} 
To characterize the phase behavior of the mixture and \tcr{quantify the presence of aggregates, we detected cheese clumps using quantile thresholding and approximated them as ellipses (see Sec. \ref{sec:data_analysis} in the supplementary material for details on data analysis). Since cheese aggregation results in a stringy sauce, a phenomenon often referred to in Italian culinary jargon as ``filatura'', we quantified aggregate sizes by measuring the major axis of the corresponding ellipses, which we refer to as aggregate size. We note that other measures for characterizing aggregate (or relative) size yielded qualitatively similar results, as shown in \fig{compare_order_p} in the supplementary material. The mean aggregate size serves as the order parameter.}  

Samples with larger, well-separated clumps resulted in higher mean aggregate sizes, while smoother mixtures corresponded to lower values. \tcr{From these measurements, we constructed a phase diagram by plotting the mean aggregate size as a function of starch concentration and temperature (\fig{vary_starch}(a)). We used a colored frame to visually associate each image with its corresponding order parameter value, measured in millimeters. To enhance interpretability, the phase diagram was smoothed using a Gaussian filter at discrete data points (\fig{vary_starch}(b)) and further refined with kernel regression smoothing to obtain a continuous representation (\fig{vary_starch}(c)). We excluded the samples in the Mozzarella Phase from the colormap, as their aggregate sizes are comparable to the system size and would distort the visualization.} The final results reveal a clear quantitative impact of starch concentration: higher starch contents shift the transition to clumpier, less homogeneous mixtures to higher temperatures. Moreover, mixtures with low starch concentrations exhibited larger aggregate sizes, while higher starch contents reduced aggregate size, \tcr{and eventually prevented the occurrence of the Mozzarella Phase}. 

Overall, these experiments corroborate the culinary insight that starch in pasta water stabilizes Cacio e pepe sauce providing a quantitative picture of its role in the cheese-water mixture's phase behavior. Starch mitigates aggregation, shifts the onset of clump formation to higher temperatures, and decreases the size of aggregates, making the sauce less sensitive to temperature control errors during preparation.
 
\section{Cheese concentration tunes the emergence of protein aggregates}
\label{sec:vary_water}

Having described the sauce properties as a function of the starch, we ought to find the mixture's degree of separation by varying the respective percentages of water and cheese. To this end, we fixed the percentage of starch in the water to $1$ \%, a value potentially enabling the appearance of rich phase behavior. Following the procedure outlined above, we performed experiments at different percentages of starch-enriched water with respect to cheese. In \fig{vary_water}(a), we show the resulting images, implementing a colored frame to highlight the mean size, namely the parameter employed to quantify the degree of separation.
As expected, by fixing the concentration of starch to 1 \%, we avoid the Mozzarella Phase emerging in some of the previous experiments (see \fig{vary_starch}). \tcr{The smoothed phase diagram in \fig{vary_water}(b) hints at the presence of a binodal region separating the homogeneous mixture from a domain where larger aggregates appear. Since proteins drive clump formation, it is natural to express the phase diagram as a function of their mass fraction, $\phi$, which we estimate from the cheese concentration using the nutritional values of Pecorino Romano DOP (see Table \ref{tab:nutritional} in the supplementary material). Applying a kernel regression smoothing finally leads to a continuous phase diagram as a function of $\phi$ and $T$ (see \fig{vary_water}(c)), which will also prove useful in the following modeling section. 
We then define the binodal line separating the aggregate-rich domain from the homogeneous one as the isoline of mean aggregate size, taken as the midpoint between the maximum and minimum observed values after smoothing.} The shape of this binodal is a paraboloid that can be fitted with a simple quadratic functional form. The minimum of the parabola 
lies slightly below the protein mass fraction value $0.134$, achieved with water and cheese in 1:1 proportion.
A parabola with positive curvature signals that both lower and higher values of protein mass fraction correspond to well-mixed sauces even at high temperatures. \tcr{Finally, we report that in the phase space lying above the binodal (see Figs.~\ref{fig:vary_starch} and~\ref{fig:vary_water}) the clumps are dynamically stable and seem not to coarsen. This observation might suggest a Turing-like mechanism, where clump formation is governed by diffusion and active chemical reactions (i.e., reactions maintained out of equilibrium by continuous replenishment of chemical fuel)~\cite{bauermannCriticalTransitionIntensive2024}. 
However, no such reactions take place in our emulsions. Instead, we believe that the lack of coarsening during experiments is simply due to the extremely slow ripening dynamics~\cite{brayTheoryPhaseorderingKinetics1994}}


\begin{figure}[t] 
\includegraphics[width=1\linewidth]{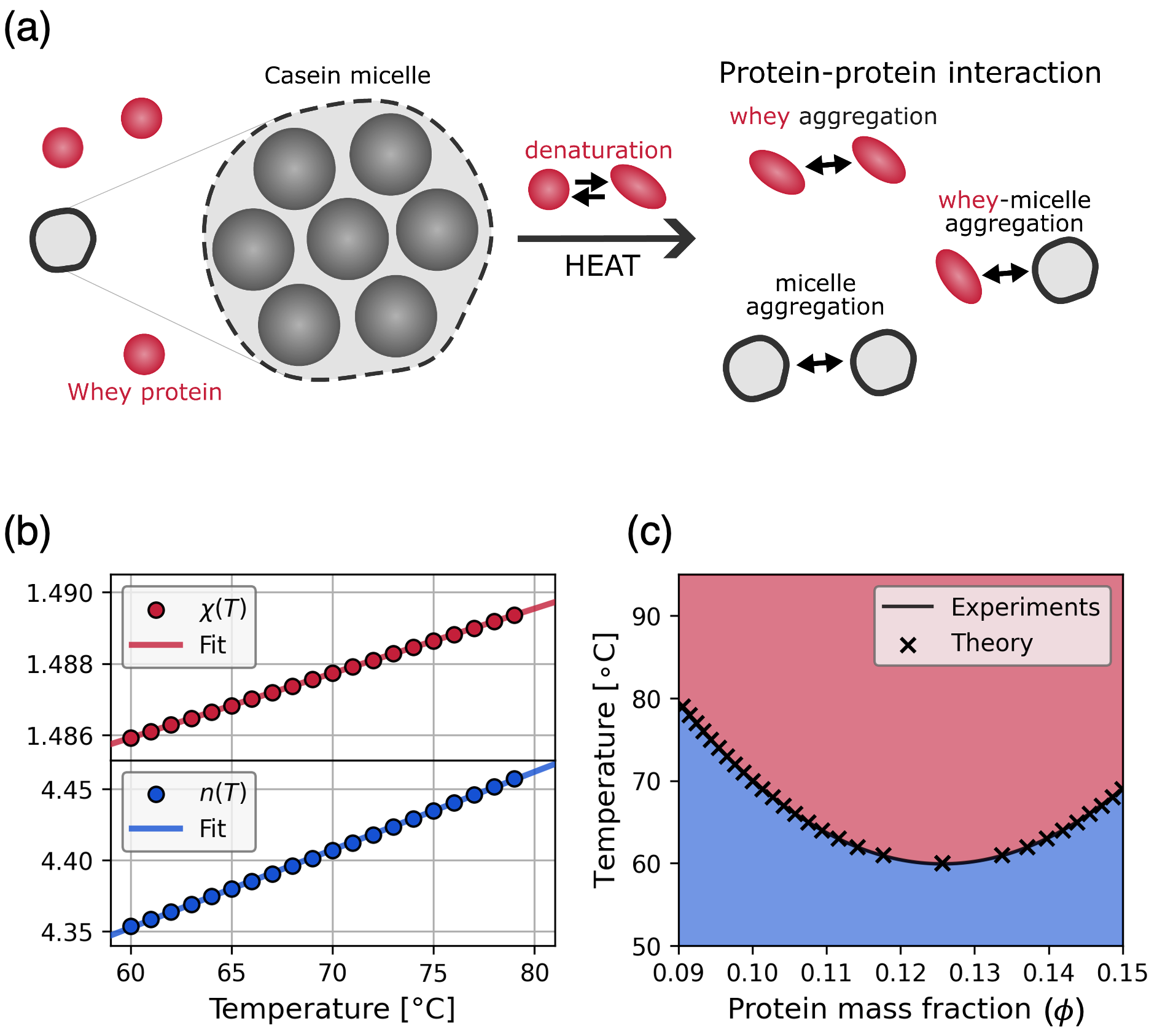}
\caption{\textbf{A minimal model for the phase behavior of Cacio e pepe sauce.} (a) Cheese is composed of casein organized in micelles together with calcium phosphate and a small percentage of whey proteins. Upon heating, whey proteins denaturate, reaching a state that favors whey-whey and whey-casein aggregation. Furthermore, casein micelles aggregate on heating. (b) Interaction and relative size (respectively $\chi$ and $n$ in the effective free energy) obtained from experiments by solving Eq.~\eqref{eq:binodal} for different temperatures. (c) Comparison between theoretical and experimental binodal curve.}
\label{fig:model}
\end{figure}

\section{A minimal model recapitulates the mixture phase behavior}
\label{sec:model}
Here, we introduce a minimal model that qualitatively matches the behavior of the mixture when the starch concentration is kept constant. Choosing which components to explicitly include in the theoretical description is a delicate step. This is because our system contains starch, salt, and lipids, together with two different kinds of proteins in the cheese, namely casein and whey (see also \fig{model}(a)). Furthermore, in the range of temperatures explored, a fraction of whey proteins undergoes denaturation \cite{donovan1987thermal,zhang2021heat}. In the following, we assume that the shape of the binodal line in the phase diagram in \fig{vary_water}(c) can be described by the phase separation of a binary mixture. We chose the binary mixture framework for its simplicity, but the applicability of such an approach has surely many limitations. In the final discussion, we outline more realistic modeling approaches and their drawbacks.

As components of our model, \tcr{as mentioned in the previous section}, we chose the cheese proteins (whey and casein) and an effective solvent encompassing water, starch, salt, and lipids constituting the cheese. We indicate the mass fractions of proteins and solvent with $\phi$, and $\phi_{\text{s}}$, respectively. Mass conservation implies $\phi + \phi_{\text{s}} = 1$. If both the components have almost equal densities, the phase diagram of a mixture in terms of the mass fraction of their components can be derived from the following free energy density~\cite{floryThermodynamicsHighPolymer1942,hugginsThermodynamicPropertiesSolutions1942}
\begin{align}
\label{eq:f}
  f = \frac{k_{\text{B}} T}{\nu} \Big[ \frac{\phi}{n} \, \ln \phi  +  (1-\phi)  \ln (&1-\phi) + \chi \phi\, (1-\phi)
    \Big]\, , \nonumber
\end{align}

where $\nu$ is the reference molecular mass, $n$ indicates the relative size of cheese proteins with respect to the other components in solutions.
To assess the phase behavior of the mixture, we introduce the exchange chemical potential $\mu = n \nu \, \partial{f}/\partial{\phi}$ and the osmotic pressure, $\Pi = -f + \phi  \,\partial{f}/\partial{\phi}$. The conditions for two phases, labelled as $\text{I}$ and $\text{II}$, to stably coexists read \cite{safranStatisticalThermodynamicsSurfaces2019}
\begin{equation}
\label{eq:binodal}
 \mu^\text{I} = \mu^\text{II} \,, \qquad \Pi^\text{I} = \Pi^\text{II}
\end{equation}
These equations have to be simultaneously solved at each temperature to determine the protein mass fraction in the coexisting phases, $\phi^\text{I}$ and $\phi^\text{II}$. As a function of temperature, $\phi^\text{I}$ and $\phi^\text{II}$ span the binodal line observed in \fig{vary_water}(c). Within the framework of binary mixtures, the minimum of the paraboloid-shaped binodal corresponds to a lower critical solution temperature. 
At this critical temperature, slightly above $60 \dc$, the two solutions collapse, i.e. $\phi^\text{I}=\phi^\text{II}$. 
In principle, one has to find the right parameters $n$ and $\chi$ so that the binodal curve we observe is duly reconstructed by this minimal model. 
Notice that inferring the dependence of these parameters on temperature can be challenging.
For this reason, we employ a reverse-engineering procedure, in analogy with \cite{fritschLocalThermodynamicsGovern2021}. We analytically solve Eq.~\eqref{eq:binodal} for $\chi$ and $n$ for each value of the temperature $T$, obtaining:
\begin{equation} \label{chi_and_n}
    \chi(T) = \mathcal{F}(\phi^\text{I},\phi^\text{II}) \qquad n(T) = \mathcal{G}(\phi^\text{I},\phi^\text{II})\,,
\end{equation}
where $\mathcal{F}$ and $\mathcal{G}$ are specified in Eq. \eqref{chi_and_n_explicit} of the supplementary material. Here, $\phi^\text{I}$ and $\phi^\text{II}$ can be directly extracted, for each value of $T$, from the experimental binodal in \fig{vary_water}(c). 
The resulting dependence on temperature of $\chi(T)$ (in red) and $n(T)$ (in blue) is reported in \fig{model}(b) and constitutes the only solution for these two parameters compatible with experiments. From a microscopic point of view, an increase in the interaction parameter $\chi$ can be due to the fact that heat induces denaturation of whey proteins, with consequential aggregation, while simultaneously favoring whey-micelle and micelle-micelle interactions \cite{WheyMicelles1,WheyMicelles2,anema2021heat} (see \fig{model}(a)). On the other hand, casein micelles are relatively heat stable, undergoing negligible dissociation on heating \cite{Micelle1}. On the same line, the increase in the relative size might be associated with whey denaturation, or it might be due to the fact that micelle and whey proteins can form starch-mediated complexes that have a larger size \cite{Complex1}. Furthermore, in \fig{model}(c) we reconstruct the binodal line employing a direct approach to Eq.~\eqref{eq:binodal} using the linear fits of $\chi(T)$ and $n(T)$ as parameters. Finally, notice that starch polymers are roughly $100$ times larger than proteins while being $100$ times less abundant in solution. As a consequence, we expected a relative size of the order of (or slightly greater than) the unity. This expectation is confirmed by our analysis, hinting at the fact that our model, despite its simplicity, is able to capture the main features of the observed phase behaviour. 

\begin{figure}[t] 
\includegraphics[width=\linewidth]{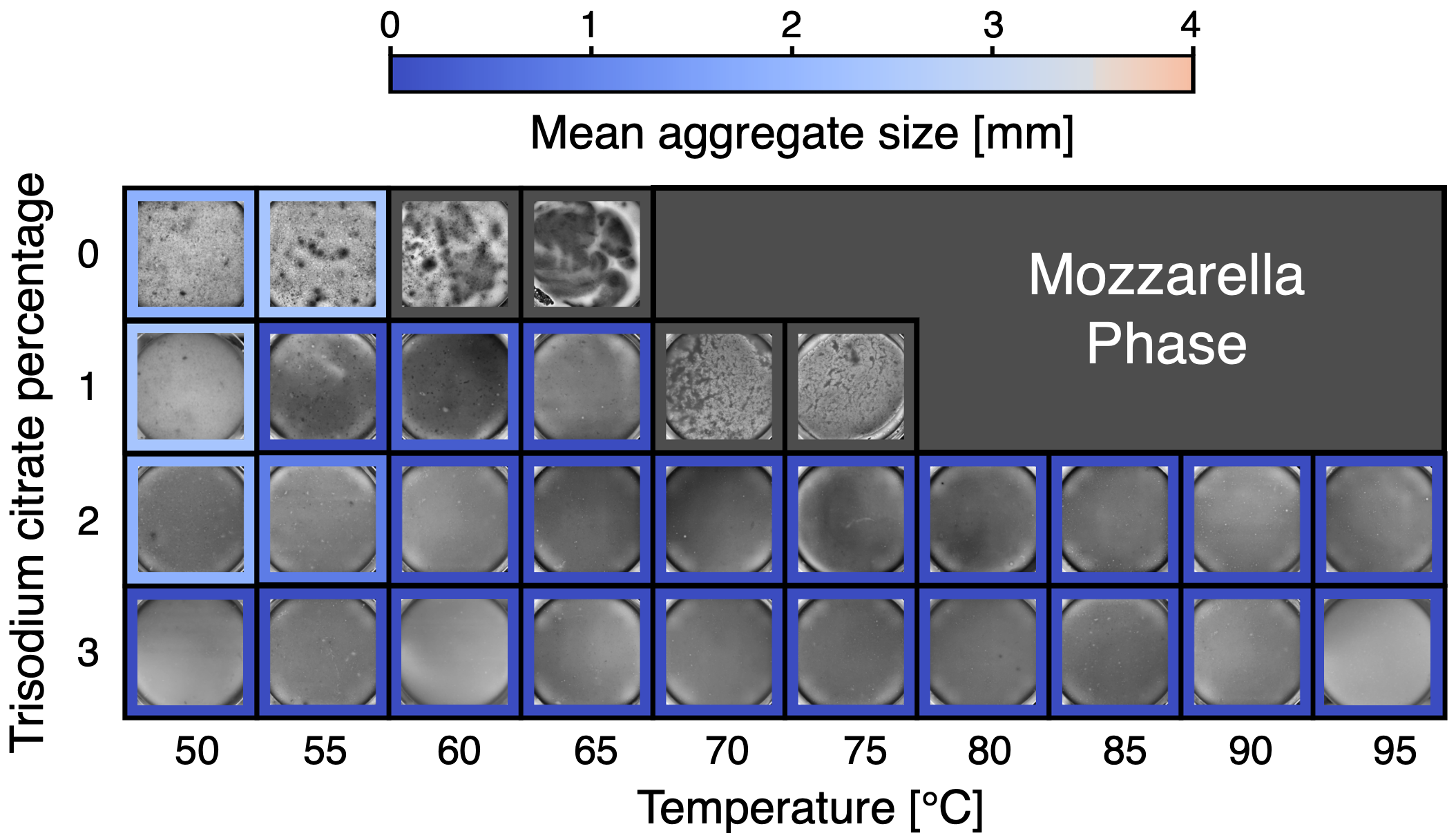}
\caption{\textbf{Effect of trisodium citrate on the stability of Cacio e pepe sauce.} The diagram shows the behavior of the sauce as a function of temperature for different concentrations of sodium citrate (1\%, 2\%, and 3\% relative to cheese mass) at a fixed 1:1 cheese-to-water ratio. Below 2\% citrate, the sauce remains unstable, leading to system-wide aggregation characteristic of the Mozzarella Phase. For concentrations of 2\% and above, the sauce is fully stabilized, preventing any visible aggregation. 
}
\label{fig:citrate}
\end{figure}

\section{Trisodium Citrate as an Alternative Stabilizer}\label{sec:citrate}

\tcr{
In this section, we investigate the effects of trisodium citrate (\textrm{Na$_3$C$_6$H$_5$O$_7$}) on the stability of the sauce. Trisodium citrate is widely used in the food industry as an emulsifier and pH buffer. Its ability to chelate calcium ions and alter protein interactions makes it particularly effective in stabilizing emulsions and improving the texture of dairy-based sauces. In particular, its chelating action primarily affects casein proteins, as calcium plays a key role in their aggregation. It is commonly used in processed cheese production to enhance smoothness and prevent undesired aggregation.}

\tcr{To quantify its effect in the context of Cacio e pepe, we applied our experimental protocol to a mixture of cheese and water at a 1:1 ratio, adding 1\%, 2\%, and 3\% of trisodium citrate relative to cheese mass. The results of our experiments are shown in \fig{citrate}. We observe that at concentrations below 2\%, the citrate is insufficient to stabilize the sauce, leading to the formation of system-wide clumps characteristic of the Mozzarella Phase, albeit at higher temperatures compared to the sauce without any stabilizer (see \fig{citrate}, second vs. first column). While these clumps qualitatively differ from those in the absence of citrate in terms of consistency, they still span the entire system.}

\tcr{For concentrations of 2\% and above, however, the sauce remains highly stable, with no visible aggregates of any size forming. This highlights that the transition to a fully stable sauce is sharper with trisodium citrate compared to starch, where small aggregates persist even in the regime where the Mozzarella Phase is absent. This difference is expected, as the macroscopic behavior directly reflects the distinct molecular mechanisms at play when citrate is present. These observations suggest that the latter can serve as an effective alternative stabilizer for Cacio e pepe, offering a reliable way to control sauce consistency, though at the cost of deviating from strict culinary tradition. However, while the sauce stabilization is more efficient, we found the taste of the cheese to be slightly blunted, likely due to the basic properties of the salt.}

\section{Scientific recipe}\label{sec:recipe}

A true Italian grandmother or a skilled home chef from Rome would never need a scientific recipe for Cacio e pepe, relying instead on instinct and years of experience. For everyone else, this guide offers a practical way to master the dish. Preparing Cacio e pepe successfully depends on getting the balance just right, particularly the ratio of starch to cheese. The concentration of starch plays a crucial role in keeping the sauce creamy and smooth, without clumps or separation. If the starch content is less than 1\% of the cheese weight, the sauce is prone to separating into unpleasant system-sized clumps corresponding to the Mozzarella Phase in  Figs.~\ref{fig:pasta} and~\ref{fig:vary_starch}. On the other hand, exceeding 4\% of starch results in a sauce that becomes stiff and unappetizing as it cools. The ideal range, as confirmed by both taste and texture tests, lies between 2\% and 3\%, ensuring stability and a pleasant consistency.

For a practical example, consider preparing Cacio e pepe for two hungry people. This typically requires 300g of pasta (tonnarelli is preferred, though spaghetti or rigatoni also works well) and 200g of cheese. \tcr{The amount of cheese can, of course, vary depending on personal taste}. Traditionalists would insist on using only \tcr{Pecorino Romano DOP}, but some argue that up to 30\% \tcr{Parmigiano Reggiano DOP} is acceptable, though this remains a point of debate. To achieve the correct starch ratio, 5g of starch is optimal for 200g of cheese.

The pasta water alone does not contain enough starch to stabilize the sauce effectively. As we already discussed, one could use pasta water ``risottata'', i.e., boiled down to concentrate the starch, but the process offers little control over the final starch amount. A more precise and reliable method is to dissolve 5g of powdered starch (such as potato or corn starch) in 50g of water. Heat this mixture gently until it thickens and turns from cloudy to nearly clear. This transition, known as starch gelatinization, \tcr{is characterized by a sudden increase in viscosity}. The next step is to combine the starch gel with the cheese. Manually grating the cheese is not ideal, since it may lead to chunks of different sizes. We recommend blending it with the starch solution for a smooth, homogeneous sauce. \tcr{To ease blending and also cool the starch gel, add 100g of water to it before adding the cheese. This increases the water-to-cheese ratio to 75\%, where, as suggested by \fig{vary_water}, the sauce is even more stable than the sauce with the same amount of starch but at 100\% water content shown in \fig{vary_starch}. Finish the sauce by adding the black pepper to the mixture. To enhance its flavors and aromas, we recommend toasting it in a pan beforehand.} 

\tcr{Alternatively, one could replace starch with 5g of trisodium citrate (2.5\% of cheese mass, dissolved in 150g of water), which provides excellent stabilization properties. However, this substitution comes at the cost of tradition, as gelatinized starch is technically an ingredient of the original recipe, while citrate is not. Also, as discussed in the previous section, 
while citrate stabilizes the sauce more effectively, it slightly blunts the cheese’s flavor, likely because of its basic properties, and changes the mouthfeel of the sauce, though not necessarily in a negative way}.

Meanwhile, cook the pasta in slightly salted water until it is al dente. Save some of the pasta cooking water before draining. Once the pasta has been drained, let it cool down for up to a minute (even a little bit longer for an amount of pasta $\ge$1kg) to prevent the excessive heat from destabilizing the sauce.  
\tcr{Indeed, in our experiments, we gradually increased the temperature of the sample over approximately 40 minutes (from 50$^\circ$C to 95$^\circ$C) to minimize kinetic effects, leading to the results shown in \fig{vary_starch} and \fig{vary_water}. While this level of control is impractical in a kitchen setting, it is important to limit thermal shocks as much as possible to achieve results closer to those observed in our experiments.}
Finally, mix the pasta with the sauce, ensuring even coating, and adjust the consistency by gradually adding reserved pasta water as needed.

One of the benefits of this stabilized sauce is its ability to withstand reheating, \tcr{most effectively in a pan}. Unlike traditional methods that risk clumping or separation, this sauce maintains its texture and stability even when brought to temperatures in the order of $80-90 \dc$. This ensures the dish can be served hot, allowing diners to enjoy it at its best. \tcr{Garnish with grated cheese and pepper before serving}.

This method offers a simple yet precise way to consistently achieve a perfect Cacio e pepe. \tcr{It is particularly useful for cooking large batches of pasta, where heat control can be challenging and requires extra care}. This recipe is inspired by Luciano Monosilio’s YouTube video \cite{YT_video}, though it does not include olive oil as suggested in his version. Despite this difference, both recipes share a focus on respecting tradition while ensuring a reliable and enjoyable result.

\section{Discussion}

In this work, we investigated the phase behavior of Cacio e pepe, one of the most famous and complicated pasta recipes of Italian culinary tradition. 
We quantified the stabilizing role of starch when mixed with cheese and water, which favors the homogeneity of the mixture up to temperatures in the order of the water boiling point. Also, by inspecting the role of cheese protein concentration, we unveil an unforeseen binodal curve resembling that of a phase-separating system with a lower critical solution temperature. By employing a minimal model to capture this phenomenology, we were able to extrapolate how protein-solution interaction and relative size behave as a function of temperature. Although it is difficult to map exactly our observations to the microscopics of this complex system, we rationalized these effective phenomena, shedding light on the interplay between cheese protein and starch under heating conditions, a widespread scenario in culinary experiences. \tcr{Additionally, we explored the stabilizing effect of trisodium citrate, which, unlike starch, induces a sharp transition from a phase-separated sauce to a homogeneous emulsion by chelating calcium ions and preventing protein aggregation}. Ultimately, our approach leads to the formulation of a scientific recipe for Cacio e pepe that capitalizes on our findings and highlights their applicative perspective.

A potential future direction could be to better understand the starch-dependent morphology of the cheese clumps. As observed in Figs.~\ref{fig:pasta} and~\ref{fig:vary_starch}, increasing the starch concentration causes the cheese-rich phase to abruptly switch from a large, system-spanning clump (the ``Mozzarella Phase'') to many smaller aggregates. This phenomenon could be explained using a model that explicitly accounts for at least three distinct species. The simplest description would account for proteins in two states A and B with stronger and weaker interaction propensity, respectively, and an effective solvent encompassing water and all remaining molecules. Then one would need to account for transitions between the two states of the proteins, controlled by temperature and starch amount. In the spirit of Ref.~\cite{bartolucciControllingCompositionCoexisting2021}, the mozzarella-like clump could represent a phase where A proteins are dominant, while small aggregates could arise whenever B proteins drive phase separation. From a microscopic point of view, A proteins could correspond to denatured whey proteins, while B proteins could encompass folded whey proteins and unfolded ones sequestered by the starch. One could even include casein micelles explicitly as a fourth component, instead of incorporating them in the effective solvent.
An even more detailed model could consider protein aggregates of different sizes explicitly. One could then couple phase separation and gelation, describing the switch to a Mozzarella Phase as the onset of a gel phase \cite{israelachviliIntermolecularSurfaceForces2015,deviriEquilibriumSizeDistribution2020,Bartolucci2024}. Such generalizations of our model could shed more light on the molecular features of this system, eventually obtaining novel insights not only into perfecting the recipe but also in the broad field of food science. However, the main issue with these multi-component approaches is that they require more experiments to estimate the dependencies of all interaction and size parameters on temperature. For example, it would be necessary to quantify the fraction of denatured whey protein for each temperature in each phase, which is a hard task. 

Other interesting future directions could include a more in-depth analysis of how starch affects effective parameters and influences the viscosity of Cacio e pepe sauce, as well as the potential role of pepper grains, the other key ingredient of the recipe, as aggregation nuclei. We hope this paper has sparked the idea that a genuine passion for fine cuisine can be translated into insightful scientific investigations, refining complex preparations and making them more accessible with everyday kitchen tools.

\section*{Supplementary Material}
The supplementary material contains: i) a detailed description of equipment, ingredients and experimental protocol employed in this study, along with a schematic representation and the actual picture of the experimental setup; ii) \tcr{a discussion of the data analysis methodology, including three examples of segmented images used to quantify the degree of separation, along with a comparison of different possible measures for assessing the mixture's phase behavior;} iii) two analytical formulas representing the expressions of $\chi$ and $n$ of the model (see Eq.~\eqref{chi_and_n}).

\section*{Acknowledgments}
\noindent
The authors thank Frank J\"ulicher, for supporting the initiative and stimulating discussions. 
We thank Tetsuya Spippayashi for enlightening clarifications on the historical origins of Cacio e pepe and Giuseppe Ricchitelli for helping with the construction of the experimental apparatus. We further thank Martina Gaiba, Alessandro Gaiba, John D. Treado, Virginia Lepore, Eleonora Nanu, Julia Kirsch, Lara Koehler, Burak Budanur, 
Irina Pi-Jaum\`a, Elizabeth Br\"uckner, M.J. Franco O\~{n}ate, Giorgio Nicoletti, and Marco Salvalaglio for their support and for eating up the sample leftovers. Finally, we thank Simone Frau for taking the photograph in \fig{pasta}(a).

\section*{Data availability}
\noindent
The data that support the findings of this study are available from the corresponding author upon reasonable request.

\bibliography{cacio&paper}

\begin{thebibliography}{51}%
\makeatletter
\providecommand \@ifxundefined [1]{%
 \@ifx{#1\undefined}
}%
\providecommand \@ifnum [1]{%
 \ifnum #1\expandafter \@firstoftwo
 \else \expandafter \@secondoftwo
 \fi
}%
\providecommand \@ifx [1]{%
 \ifx #1\expandafter \@firstoftwo
 \else \expandafter \@secondoftwo
 \fi
}%
\providecommand \natexlab [1]{#1}%
\providecommand \enquote  [1]{``#1''}%
\providecommand \bibnamefont  [1]{#1}%
\providecommand \bibfnamefont [1]{#1}%
\providecommand \citenamefont [1]{#1}%
\providecommand \href@noop [0]{\@secondoftwo}%
\providecommand \href [0]{\begingroup \@sanitize@url \@href}%
\providecommand \@href[1]{\@@startlink{#1}\@@href}%
\providecommand \@@href[1]{\endgroup#1\@@endlink}%
\providecommand \@sanitize@url [0]{\catcode `\\12\catcode `\$12\catcode `\&12\catcode `\#12\catcode `\^12\catcode `\_12\catcode `\%12\relax}%
\providecommand \@@startlink[1]{}%
\providecommand \@@endlink[0]{}%
\providecommand \url  [0]{\begingroup\@sanitize@url \@url }%
\providecommand \@url [1]{\endgroup\@href {#1}{\urlprefix }}%
\providecommand \urlprefix  [0]{URL }%
\providecommand \Eprint [0]{\href }%
\providecommand \doibase [0]{https://doi.org/}%
\providecommand \selectlanguage [0]{\@gobble}%
\providecommand \bibinfo  [0]{\@secondoftwo}%
\providecommand \bibfield  [0]{\@secondoftwo}%
\providecommand \translation [1]{[#1]}%
\providecommand \BibitemOpen [0]{}%
\providecommand \bibitemStop [0]{}%
\providecommand \bibitemNoStop [0]{.\EOS\space}%
\providecommand \EOS [0]{\spacefactor3000\relax}%
\providecommand \BibitemShut  [1]{\csname bibitem#1\endcsname}%
\let\auto@bib@innerbib\@empty
\bibitem [{\citenamefont {Ball}(2010)}]{ballPastaPhysics2010}%
  \BibitemOpen
  \bibfield  {author} {\bibinfo {author} {\bibfnamefont {P.}~\bibnamefont {Ball}},\ }\bibfield  {title} {\bibinfo {title} {Pasta physics},\ }\href {https://doi.org/10.1038/nmat2793} {\bibfield  {journal} {\bibinfo  {journal} {Nature Materials}\ }\textbf {\bibinfo {volume} {9}},\ \bibinfo {pages} {539} (\bibinfo {year} {2010})}\BibitemShut {NoStop}%
\bibitem [{\citenamefont {Audoly}(2005)}]{audolyFragmentationRodsCascading2005}%
  \BibitemOpen
  \bibfield  {author} {\bibinfo {author} {\bibfnamefont {B.}~\bibnamefont {Audoly}},\ }\bibfield  {title} {\bibinfo {title} {Fragmentation of {{Rods}} by {{Cascading Cracks}}: {{Why Spaghetti Does Not Break}} in {{Half}}},\ }\bibfield  {journal} {\bibinfo  {journal} {Physical Review Letters}\ }\textbf {\bibinfo {volume} {95}},\ \href {https://doi.org/10.1103/PhysRevLett.95.095505} {10.1103/PhysRevLett.95.095505} (\bibinfo {year} {2005})\BibitemShut {NoStop}%
\bibitem [{\citenamefont {Caulkin}\ \emph {et~al.}(2010)\citenamefont {Caulkin}, \citenamefont {Jia}, \citenamefont {Fairweather},\ and\ \citenamefont {Williams}}]{caulkinGeometricAspectsParticle2010}%
  \BibitemOpen
  \bibfield  {author} {\bibinfo {author} {\bibfnamefont {R.}~\bibnamefont {Caulkin}}, \bibinfo {author} {\bibfnamefont {X.}~\bibnamefont {Jia}}, \bibinfo {author} {\bibfnamefont {M.}~\bibnamefont {Fairweather}},\ and\ \bibinfo {author} {\bibfnamefont {R.~A.}\ \bibnamefont {Williams}},\ }\bibfield  {title} {\bibinfo {title} {Geometric aspects of particle segregation},\ }\href {https://doi.org/10.1103/PhysRevE.81.051302} {\bibfield  {journal} {\bibinfo  {journal} {Physical Review E}\ }\textbf {\bibinfo {volume} {81}},\ \bibinfo {pages} {051302} (\bibinfo {year} {2010})}\BibitemShut {NoStop}%
\bibitem [{\citenamefont {Tao}\ \emph {et~al.}(2021)\citenamefont {Tao}, \citenamefont {Lee}, \citenamefont {Liu}, \citenamefont {Zhang}, \citenamefont {Cui}, \citenamefont {Mondoa}, \citenamefont {Babaei}, \citenamefont {Santillan}, \citenamefont {Wang}, \citenamefont {Luo}, \citenamefont {Liu}, \citenamefont {Yang}, \citenamefont {Do}, \citenamefont {Sun}, \citenamefont {Wang}, \citenamefont {Zhang},\ and\ \citenamefont {Yao}}]{taoMorphingPasta2021}%
  \BibitemOpen
  \bibfield  {author} {\bibinfo {author} {\bibfnamefont {Y.}~\bibnamefont {Tao}}, \bibinfo {author} {\bibfnamefont {Y.-C.}\ \bibnamefont {Lee}}, \bibinfo {author} {\bibfnamefont {H.}~\bibnamefont {Liu}}, \bibinfo {author} {\bibfnamefont {X.}~\bibnamefont {Zhang}}, \bibinfo {author} {\bibfnamefont {J.}~\bibnamefont {Cui}}, \bibinfo {author} {\bibfnamefont {C.}~\bibnamefont {Mondoa}}, \bibinfo {author} {\bibfnamefont {M.}~\bibnamefont {Babaei}}, \bibinfo {author} {\bibfnamefont {J.}~\bibnamefont {Santillan}}, \bibinfo {author} {\bibfnamefont {G.}~\bibnamefont {Wang}}, \bibinfo {author} {\bibfnamefont {D.}~\bibnamefont {Luo}}, \bibinfo {author} {\bibfnamefont {D.}~\bibnamefont {Liu}}, \bibinfo {author} {\bibfnamefont {H.}~\bibnamefont {Yang}}, \bibinfo {author} {\bibfnamefont {Y.}~\bibnamefont {Do}}, \bibinfo {author} {\bibfnamefont {L.}~\bibnamefont {Sun}}, \bibinfo {author} {\bibfnamefont {W.}~\bibnamefont {Wang}}, \bibinfo {author} {\bibfnamefont {T.}~\bibnamefont {Zhang}},\ and\ \bibinfo {author}
  {\bibfnamefont {L.}~\bibnamefont {Yao}},\ }\bibfield  {title} {\bibinfo {title} {Morphing pasta and beyond},\ }\href {https://doi.org/10.1126/sciadv.abf4098} {\bibfield  {journal} {\bibinfo  {journal} {Science Advances}\ }\textbf {\bibinfo {volume} {7}},\ \bibinfo {pages} {eabf4098} (\bibinfo {year} {2021})}\BibitemShut {NoStop}%
\bibitem [{\citenamefont {Hwang}\ \emph {et~al.}(2022)\citenamefont {Hwang}, \citenamefont {Ha}, \citenamefont {Siu}, \citenamefont {Kim},\ and\ \citenamefont {Tawfick}}]{hwangSwellingSofteningElastocapillary2022}%
  \BibitemOpen
  \bibfield  {author} {\bibinfo {author} {\bibfnamefont {J.}~\bibnamefont {Hwang}}, \bibinfo {author} {\bibfnamefont {J.}~\bibnamefont {Ha}}, \bibinfo {author} {\bibfnamefont {R.}~\bibnamefont {Siu}}, \bibinfo {author} {\bibfnamefont {Y.~S.}\ \bibnamefont {Kim}},\ and\ \bibinfo {author} {\bibfnamefont {S.}~\bibnamefont {Tawfick}},\ }\bibfield  {title} {\bibinfo {title} {Swelling, softening, and elastocapillary adhesion of cooked pasta},\ }\href {https://doi.org/10.1063/5.0083696} {\bibfield  {journal} {\bibinfo  {journal} {Physics of Fluids}\ }\textbf {\bibinfo {volume} {34}},\ \bibinfo {pages} {042105} (\bibinfo {year} {2022})}\BibitemShut {NoStop}%
\bibitem [{\citenamefont {Goldberg}\ and\ \citenamefont {O'Reilly}(2020)}]{goldbergMechanicsbasedModelCookinginduced2020}%
  \BibitemOpen
  \bibfield  {author} {\bibinfo {author} {\bibfnamefont {N.~N.}\ \bibnamefont {Goldberg}}\ and\ \bibinfo {author} {\bibfnamefont {O.~M.}\ \bibnamefont {O'Reilly}},\ }\bibfield  {title} {\bibinfo {title} {Mechanics-based model for the cooking-induced deformation of spaghetti},\ }\href {https://doi.org/10.1103/PhysRevE.101.013001} {\bibfield  {journal} {\bibinfo  {journal} {Physical Review E}\ }\textbf {\bibinfo {volume} {101}},\ \bibinfo {pages} {013001} (\bibinfo {year} {2020})}\BibitemShut {NoStop}%
\bibitem [{\citenamefont {Gonzalez}\ \emph {et~al.}(2000)\citenamefont {Gonzalez}, \citenamefont {Mccarthy},\ and\ \citenamefont {Mccarthy}}]{gonzalezTexturalStructuralChanges2000}%
  \BibitemOpen
  \bibfield  {author} {\bibinfo {author} {\bibfnamefont {J.~J.}\ \bibnamefont {Gonzalez}}, \bibinfo {author} {\bibfnamefont {K.~L.}\ \bibnamefont {Mccarthy}},\ and\ \bibinfo {author} {\bibfnamefont {M.~J.}\ \bibnamefont {Mccarthy}},\ }\bibfield  {title} {\bibinfo {title} {Textural and {{Structural Changes}} in {{Lasagna After Cooking}}},\ }\href {https://doi.org/10.1111/j.1745-4603.2000.tb00286.x} {\bibfield  {journal} {\bibinfo  {journal} {Journal of Texture Studies}\ }\textbf {\bibinfo {volume} {31}},\ \bibinfo {pages} {93} (\bibinfo {year} {2000})}\BibitemShut {NoStop}%
\bibitem [{\citenamefont {Michieletto}\ and\ \citenamefont {Turner}(2014)}]{michielettoTasteAnelloni2014}%
  \BibitemOpen
  \bibfield  {author} {\bibinfo {author} {\bibfnamefont {D.}~\bibnamefont {Michieletto}}\ and\ \bibinfo {author} {\bibfnamefont {M.~S.}\ \bibnamefont {Turner}},\ }\bibfield  {title} {\bibinfo {title} {A taste for anelloni},\ }\href {https://doi.org/10.1088/2058-7058/27/12/36} {\bibfield  {journal} {\bibinfo  {journal} {Physics World}\ }\textbf {\bibinfo {volume} {27}},\ \bibinfo {pages} {28} (\bibinfo {year} {2014})}\BibitemShut {NoStop}%
\bibitem [{\citenamefont {Caplan}\ and\ \citenamefont {Horowitz}(2017)}]{caplanColloquiumAstromaterialScience2017}%
  \BibitemOpen
  \bibfield  {author} {\bibinfo {author} {\bibfnamefont {M.~E.}\ \bibnamefont {Caplan}}\ and\ \bibinfo {author} {\bibfnamefont {C.~J.}\ \bibnamefont {Horowitz}},\ }\bibfield  {title} {\bibinfo {title} {Colloquium: {{Astromaterial}} science and nuclear pasta},\ }\href {https://doi.org/10.1103/RevModPhys.89.041002} {\bibfield  {journal} {\bibinfo  {journal} {Reviews of Modern Physics}\ }\textbf {\bibinfo {volume} {89}},\ \bibinfo {pages} {041002} (\bibinfo {year} {2017})}\BibitemShut {NoStop}%
\bibitem [{\citenamefont {Jenkins}\ and\ \citenamefont {Donald}(1998)}]{jenkinsGelatinisationStarchCombined1998}%
  \BibitemOpen
  \bibfield  {author} {\bibinfo {author} {\bibfnamefont {P.~J.}\ \bibnamefont {Jenkins}}\ and\ \bibinfo {author} {\bibfnamefont {A.~M.}\ \bibnamefont {Donald}},\ }\bibfield  {title} {\bibinfo {title} {Gelatinisation of starch: A combined {{SAXS}}/{{WAXS}}/{{DSC}} and {{SANS}} study},\ }\href {https://doi.org/10.1016/S0008-6215(98)00079-2} {\bibfield  {journal} {\bibinfo  {journal} {Carbohydrate Research}\ }\textbf {\bibinfo {volume} {308}},\ \bibinfo {pages} {133} (\bibinfo {year} {1998})}\BibitemShut {NoStop}%
\bibitem [{\citenamefont {Tako}\ \emph {et~al.}(2014)\citenamefont {Tako}, \citenamefont {Tamaki}, \citenamefont {Teruya},\ and\ \citenamefont {Takeda}}]{takoPrinciplesStarchGelatinization2014}%
  \BibitemOpen
  \bibfield  {author} {\bibinfo {author} {\bibfnamefont {M.}~\bibnamefont {Tako}}, \bibinfo {author} {\bibfnamefont {Y.}~\bibnamefont {Tamaki}}, \bibinfo {author} {\bibfnamefont {T.}~\bibnamefont {Teruya}},\ and\ \bibinfo {author} {\bibfnamefont {Y.}~\bibnamefont {Takeda}},\ }\bibfield  {title} {\bibinfo {title} {The {{Principles}} of {{Starch Gelatinization}} and {{Retrogradation}}},\ }\href {https://doi.org/10.4236/fns.2014.53035} {\bibfield  {journal} {\bibinfo  {journal} {Food and Nutrition Sciences}\ }\textbf {\bibinfo {volume} {05}},\ \bibinfo {pages} {280} (\bibinfo {year} {2014})}\BibitemShut {NoStop}%
\bibitem [{\citenamefont {Krishna}\ \emph {et~al.}(2021)\citenamefont {Krishna}, \citenamefont {Hussain}, \citenamefont {Kiran},\ and\ \citenamefont {Kumar}}]{krishnaExperimentalEvaluationImpact2021}%
  \BibitemOpen
  \bibfield  {author} {\bibinfo {author} {\bibfnamefont {V.~S.~R.}\ \bibnamefont {Krishna}}, \bibinfo {author} {\bibfnamefont {S.}~\bibnamefont {Hussain}}, \bibinfo {author} {\bibfnamefont {C.~R.}\ \bibnamefont {Kiran}},\ and\ \bibinfo {author} {\bibfnamefont {K.~V.}\ \bibnamefont {Kumar}},\ }\bibfield  {title} {\bibinfo {title} {Experimental evaluation of impact energy on oobleck material (non-{{Newtonian}} fluid)},\ }\href {https://doi.org/10.1016/j.matpr.2020.12.1112} {\bibfield  {journal} {\bibinfo  {journal} {Materials Today: Proceedings}\ }\bibinfo {series} {International {{Conference}} on {{Advances}} in {{Materials Research}} - 2019},\ \textbf {\bibinfo {volume} {45}},\ \bibinfo {pages} {3609} (\bibinfo {year} {2021})}\BibitemShut {NoStop}%
\bibitem [{\citenamefont {Mathijssen}\ \emph {et~al.}(2023)\citenamefont {Mathijssen}, \citenamefont {Lisicki}, \citenamefont {Prakash},\ and\ \citenamefont {Mossige}}]{mathijssenCulinaryFluidMechanics2023}%
  \BibitemOpen
  \bibfield  {author} {\bibinfo {author} {\bibfnamefont {A.~J. T.~M.}\ \bibnamefont {Mathijssen}}, \bibinfo {author} {\bibfnamefont {M.}~\bibnamefont {Lisicki}}, \bibinfo {author} {\bibfnamefont {V.~N.}\ \bibnamefont {Prakash}},\ and\ \bibinfo {author} {\bibfnamefont {E.~J.~L.}\ \bibnamefont {Mossige}},\ }\bibfield  {title} {\bibinfo {title} {Culinary fluid mechanics and other currents in food science},\ }\href {https://doi.org/10.1103/RevModPhys.95.025004} {\bibfield  {journal} {\bibinfo  {journal} {Reviews of Modern Physics}\ }\textbf {\bibinfo {volume} {95}},\ \bibinfo {pages} {025004} (\bibinfo {year} {2023})}\BibitemShut {NoStop}%
\bibitem [{\citenamefont {Tolstoguzov}(2006)}]{tolstoguzovTexturisingPhaseSeparation2006}%
  \BibitemOpen
  \bibfield  {author} {\bibinfo {author} {\bibfnamefont {V.}~\bibnamefont {Tolstoguzov}},\ }\bibfield  {title} {\bibinfo {title} {Texturising by phase separation},\ }\href {https://doi.org/10.1016/j.biotechadv.2006.07.001} {\bibfield  {journal} {\bibinfo  {journal} {Biotechnology Advances}\ }\textbf {\bibinfo {volume} {24}},\ \bibinfo {pages} {626} (\bibinfo {year} {2006})}\BibitemShut {NoStop}%
\bibitem [{\citenamefont {Friberg}\ \emph {et~al.}(2003)\citenamefont {Friberg}, \citenamefont {Larsson},\ and\ \citenamefont {Sjoblom}}]{fribergFoodEmulsions2003}%
  \BibitemOpen
  \bibinfo {editor} {\bibfnamefont {S.}~\bibnamefont {Friberg}}, \bibinfo {editor} {\bibfnamefont {K.}~\bibnamefont {Larsson}},\ and\ \bibinfo {editor} {\bibfnamefont {J.}~\bibnamefont {Sjoblom}},\ eds.,\ \href {https://doi.org/10.1201/9780203913222} {\emph {\bibinfo {title} {Food {{Emulsions}}}}},\ \bibinfo {edition} {4th}\ ed.\ (\bibinfo  {publisher} {CRC Press},\ \bibinfo {address} {Boca Raton},\ \bibinfo {year} {2003})\BibitemShut {NoStop}%
\bibitem [{\citenamefont {Dickinson}(2010)}]{dickinsonFoodEmulsionsFoams2010}%
  \BibitemOpen
  \bibfield  {author} {\bibinfo {author} {\bibfnamefont {E.}~\bibnamefont {Dickinson}},\ }\bibfield  {title} {\bibinfo {title} {Food emulsions and foams: {{Stabilization}} by particles},\ }\href {https://doi.org/10.1016/j.cocis.2009.11.001} {\bibfield  {journal} {\bibinfo  {journal} {Current Opinion in Colloid \& Interface Science}\ }\textbf {\bibinfo {volume} {15}},\ \bibinfo {pages} {40} (\bibinfo {year} {2010})}\BibitemShut {NoStop}%
\bibitem [{\citenamefont {Koizumi}\ \emph {et~al.}(2022)\citenamefont {Koizumi}, \citenamefont {Takagi}, \citenamefont {Hondoh}, \citenamefont {Michikawa}, \citenamefont {Hirai},\ and\ \citenamefont {Ueno}}]{koizumiControlPhaseSeparation2022}%
  \BibitemOpen
  \bibfield  {author} {\bibinfo {author} {\bibfnamefont {H.}~\bibnamefont {Koizumi}}, \bibinfo {author} {\bibfnamefont {M.}~\bibnamefont {Takagi}}, \bibinfo {author} {\bibfnamefont {H.}~\bibnamefont {Hondoh}}, \bibinfo {author} {\bibfnamefont {S.}~\bibnamefont {Michikawa}}, \bibinfo {author} {\bibfnamefont {Y.}~\bibnamefont {Hirai}},\ and\ \bibinfo {author} {\bibfnamefont {S.}~\bibnamefont {Ueno}},\ }\bibfield  {title} {\bibinfo {title} {Control of {{Phase Separation}} for {{CBS-Based Compound Chocolates Focusing}} on {{Growth Kinetics}}},\ }\href {https://doi.org/10.1021/acs.cgd.2c00317} {\bibfield  {journal} {\bibinfo  {journal} {Crystal Growth \& Design}\ }\textbf {\bibinfo {volume} {22}},\ \bibinfo {pages} {6879} (\bibinfo {year} {2022})}\BibitemShut {NoStop}%
\bibitem [{\citenamefont {Aichinger}\ \emph {et~al.}(2017)\citenamefont {Aichinger}, \citenamefont {Schmitt}, \citenamefont {Gunes}, \citenamefont {Leser}, \citenamefont {Sagalowicz},\ and\ \citenamefont {Michel}}]{aichingerPhaseSeparationFood2017}%
  \BibitemOpen
  \bibfield  {author} {\bibinfo {author} {\bibfnamefont {P.-A.}\ \bibnamefont {Aichinger}}, \bibinfo {author} {\bibfnamefont {C.}~\bibnamefont {Schmitt}}, \bibinfo {author} {\bibfnamefont {D.~Z.}\ \bibnamefont {Gunes}}, \bibinfo {author} {\bibfnamefont {M.~E.}\ \bibnamefont {Leser}}, \bibinfo {author} {\bibfnamefont {L.}~\bibnamefont {Sagalowicz}},\ and\ \bibinfo {author} {\bibfnamefont {M.}~\bibnamefont {Michel}},\ }\bibfield  {title} {\bibinfo {title} {Phase separation in food material design inspired by {{Nature}}: {{Or}}: {{What}} ice cream can learn from frogs},\ }\href {https://doi.org/10.1016/j.cocis.2017.03.002} {\bibfield  {journal} {\bibinfo  {journal} {Current Opinion in Colloid \& Interface Science}\ }\textbf {\bibinfo {volume} {28}},\ \bibinfo {pages} {56} (\bibinfo {year} {2017})}\BibitemShut {NoStop}%
\bibitem [{\citenamefont {Vratsanos}\ \emph {et~al.}(2023)\citenamefont {Vratsanos}, \citenamefont {Xue}, \citenamefont {Rosenmann}, \citenamefont {Zarzar},\ and\ \citenamefont {Gianneschi}}]{vratsanosOuzoEffectExamined2023}%
  \BibitemOpen
  \bibfield  {author} {\bibinfo {author} {\bibfnamefont {M.~A.}\ \bibnamefont {Vratsanos}}, \bibinfo {author} {\bibfnamefont {W.}~\bibnamefont {Xue}}, \bibinfo {author} {\bibfnamefont {N.~D.}\ \bibnamefont {Rosenmann}}, \bibinfo {author} {\bibfnamefont {L.~D.}\ \bibnamefont {Zarzar}},\ and\ \bibinfo {author} {\bibfnamefont {N.~C.}\ \bibnamefont {Gianneschi}},\ }\bibfield  {title} {\bibinfo {title} {Ouzo {{Effect Examined}} at the {{Nanoscale}} via {{Direct Observation}} of {{Droplet Nucleation}} and {{Morphology}}},\ }\href {https://doi.org/10.1021/acscentsci.2c01194} {\bibfield  {journal} {\bibinfo  {journal} {ACS Central Science}\ }\textbf {\bibinfo {volume} {9}},\ \bibinfo {pages} {457} (\bibinfo {year} {2023})}\BibitemShut {NoStop}%
\bibitem [{\citenamefont {Grillo}(2003)}]{grilloSmallangleNeutronScattering2003}%
  \BibitemOpen
  \bibfield  {author} {\bibinfo {author} {\bibfnamefont {I.}~\bibnamefont {Grillo}},\ }\bibfield  {title} {\bibinfo {title} {Small-angle neutron scattering study of a world-wide known emulsion: {{Le Pastis}}},\ }\href {https://doi.org/10.1016/S0927-7757(03)00331-5} {\bibfield  {journal} {\bibinfo  {journal} {Colloids and Surfaces A: Physicochemical and Engineering Aspects}\ }\textbf {\bibinfo {volume} {225}},\ \bibinfo {pages} {153} (\bibinfo {year} {2003})}\BibitemShut {NoStop}%
\bibitem [{\citenamefont {Chiappisi}\ and\ \citenamefont {Grillo}(2018)}]{chiappisiLookingLimoncelloStructure2018}%
  \BibitemOpen
  \bibfield  {author} {\bibinfo {author} {\bibfnamefont {L.}~\bibnamefont {Chiappisi}}\ and\ \bibinfo {author} {\bibfnamefont {I.}~\bibnamefont {Grillo}},\ }\bibfield  {title} {\bibinfo {title} {Looking into {{Limoncello}}: {{The Structure}} of the {{Italian Liquor Revealed}} by {{Small-Angle Neutron Scattering}}},\ }\href {https://doi.org/10.1021/acsomega.8b01858} {\bibfield  {journal} {\bibinfo  {journal} {ACS Omega}\ }\textbf {\bibinfo {volume} {3}},\ \bibinfo {pages} {15407} (\bibinfo {year} {2018})}\BibitemShut {NoStop}%
\bibitem [{\citenamefont {Brangwynne}\ \emph {et~al.}(2009)\citenamefont {Brangwynne}, \citenamefont {Eckmann}, \citenamefont {Courson}, \citenamefont {Rybarska}, \citenamefont {Hoege}, \citenamefont {Gharakhani}, \citenamefont {J{\"u}licher},\ and\ \citenamefont {Hyman}}]{brangwynneGermlineGranulesAre2009}%
  \BibitemOpen
  \bibfield  {author} {\bibinfo {author} {\bibfnamefont {C.~P.}\ \bibnamefont {Brangwynne}}, \bibinfo {author} {\bibfnamefont {C.~R.}\ \bibnamefont {Eckmann}}, \bibinfo {author} {\bibfnamefont {D.~S.}\ \bibnamefont {Courson}}, \bibinfo {author} {\bibfnamefont {A.}~\bibnamefont {Rybarska}}, \bibinfo {author} {\bibfnamefont {C.}~\bibnamefont {Hoege}}, \bibinfo {author} {\bibfnamefont {J.}~\bibnamefont {Gharakhani}}, \bibinfo {author} {\bibfnamefont {F.}~\bibnamefont {J{\"u}licher}},\ and\ \bibinfo {author} {\bibfnamefont {A.~A.}\ \bibnamefont {Hyman}},\ }\bibfield  {title} {\bibinfo {title} {Germline {{P Granules Are Liquid Droplets That Localize}} by {{Controlled Dissolution}}/{{Condensation}}},\ }\href {https://doi.org/10.1126/science.1172046} {\bibfield  {journal} {\bibinfo  {journal} {Science}\ }\textbf {\bibinfo {volume} {324}},\ \bibinfo {pages} {1729} (\bibinfo {year} {2009})}\BibitemShut {NoStop}%
\bibitem [{\citenamefont {Hyman}\ \emph {et~al.}(2014)\citenamefont {Hyman}, \citenamefont {Weber},\ and\ \citenamefont {J{\"u}licher}}]{hymanLiquidLiquidPhaseSeparation2014}%
  \BibitemOpen
  \bibfield  {author} {\bibinfo {author} {\bibfnamefont {A.~A.}\ \bibnamefont {Hyman}}, \bibinfo {author} {\bibfnamefont {C.~A.}\ \bibnamefont {Weber}},\ and\ \bibinfo {author} {\bibfnamefont {F.}~\bibnamefont {J{\"u}licher}},\ }\bibfield  {title} {\bibinfo {title} {Liquid-{{Liquid Phase Separation}} in {{Biology}}},\ }\href {https://doi.org/10.1146/annurev-cellbio-100913-013325} {\bibfield  {journal} {\bibinfo  {journal} {Annual Review of Cell and Developmental Biology}\ }\textbf {\bibinfo {volume} {30}},\ \bibinfo {pages} {39} (\bibinfo {year} {2014})}\BibitemShut {NoStop}%
\bibitem [{\citenamefont {Banani}\ \emph {et~al.}(2017)\citenamefont {Banani}, \citenamefont {Lee}, \citenamefont {Hyman},\ and\ \citenamefont {Rosen}}]{bananiBiomolecularCondensatesOrganizers2017}%
  \BibitemOpen
  \bibfield  {author} {\bibinfo {author} {\bibfnamefont {S.~F.}\ \bibnamefont {Banani}}, \bibinfo {author} {\bibfnamefont {H.~O.}\ \bibnamefont {Lee}}, \bibinfo {author} {\bibfnamefont {A.~A.}\ \bibnamefont {Hyman}},\ and\ \bibinfo {author} {\bibfnamefont {M.~K.}\ \bibnamefont {Rosen}},\ }\bibfield  {title} {\bibinfo {title} {Biomolecular condensates: Organizers of cellular biochemistry},\ }\href {https://doi.org/10.1038/nrm.2017.7} {\bibfield  {journal} {\bibinfo  {journal} {Nature Reviews Molecular Cell Biology}\ }\textbf {\bibinfo {volume} {18}},\ \bibinfo {pages} {285} (\bibinfo {year} {2017})}\BibitemShut {NoStop}%
\bibitem [{\citenamefont {Alberti}\ \emph {et~al.}(2019)\citenamefont {Alberti}, \citenamefont {Gladfelter},\ and\ \citenamefont {Mittag}}]{albertiConsiderationsChallengesStudying2019}%
  \BibitemOpen
  \bibfield  {author} {\bibinfo {author} {\bibfnamefont {S.}~\bibnamefont {Alberti}}, \bibinfo {author} {\bibfnamefont {A.}~\bibnamefont {Gladfelter}},\ and\ \bibinfo {author} {\bibfnamefont {T.}~\bibnamefont {Mittag}},\ }\bibfield  {title} {\bibinfo {title} {Considerations and {{Challenges}} in {{Studying Liquid-Liquid Phase Separation}} and {{Biomolecular Condensates}}},\ }\href {https://doi.org/10.1016/j.cell.2018.12.035} {\bibfield  {journal} {\bibinfo  {journal} {Cell}\ }\textbf {\bibinfo {volume} {176}},\ \bibinfo {pages} {419} (\bibinfo {year} {2019})}\BibitemShut {NoStop}%
\bibitem [{\citenamefont {Zbinden}\ \emph {et~al.}(2020)\citenamefont {Zbinden}, \citenamefont {{P{\'e}rez-Berlanga}}, \citenamefont {De~Rossi},\ and\ \citenamefont {Polymenidou}}]{zbindenPhaseSeparationNeurodegenerative2020}%
  \BibitemOpen
  \bibfield  {author} {\bibinfo {author} {\bibfnamefont {A.}~\bibnamefont {Zbinden}}, \bibinfo {author} {\bibfnamefont {M.}~\bibnamefont {{P{\'e}rez-Berlanga}}}, \bibinfo {author} {\bibfnamefont {P.}~\bibnamefont {De~Rossi}},\ and\ \bibinfo {author} {\bibfnamefont {M.}~\bibnamefont {Polymenidou}},\ }\bibfield  {title} {\bibinfo {title} {Phase {{Separation}} and {{Neurodegenerative Diseases}}: {{A Disturbance}} in the {{Force}}},\ }\href {https://doi.org/10.1016/j.devcel.2020.09.014} {\bibfield  {journal} {\bibinfo  {journal} {Developmental Cell}\ }\textbf {\bibinfo {volume} {55}},\ \bibinfo {pages} {45} (\bibinfo {year} {2020})}\BibitemShut {NoStop}%
\bibitem [{\citenamefont {Oparin}(1952)}]{oparinOriginLife1952}%
  \BibitemOpen
  \bibfield  {author} {\bibinfo {author} {\bibfnamefont {A.~I.}\ \bibnamefont {Oparin}},\ }\href@noop {} {\emph {\bibinfo {title} {The {{Origin Of Life}}}}}\ (\bibinfo  {publisher} {Dover Publications},\ \bibinfo {year} {1952})\BibitemShut {NoStop}%
\bibitem [{\citenamefont {Haldane}(1929)}]{haldaneOriginLife1929}%
  \BibitemOpen
  \bibfield  {author} {\bibinfo {author} {\bibfnamefont {J.~B.~S.}\ \bibnamefont {Haldane}},\ }\bibfield  {title} {\bibinfo {title} {The origin of life},\ }\href@noop {} {\bibfield  {journal} {\bibinfo  {journal} {Rationalist Annual}\ }\textbf {\bibinfo {volume} {148}},\ \bibinfo {pages} {3} (\bibinfo {year} {1929})}\BibitemShut {NoStop}%
\bibitem [{\citenamefont {Morasch}\ \emph {et~al.}(2019)\citenamefont {Morasch}, \citenamefont {Liu}, \citenamefont {Dirscherl}, \citenamefont {Ianeselli}, \citenamefont {K{\"u}hnlein}, \citenamefont {Le~Vay}, \citenamefont {Schwintek}, \citenamefont {Islam}, \citenamefont {Corpinot}, \citenamefont {Scheu}, \citenamefont {Dingwell}, \citenamefont {Schwille}, \citenamefont {Mutschler}, \citenamefont {Powner}, \citenamefont {Mast},\ and\ \citenamefont {Braun}}]{moraschHeatedGasBubbles2019}%
  \BibitemOpen
  \bibfield  {author} {\bibinfo {author} {\bibfnamefont {M.}~\bibnamefont {Morasch}}, \bibinfo {author} {\bibfnamefont {J.}~\bibnamefont {Liu}}, \bibinfo {author} {\bibfnamefont {C.~F.}\ \bibnamefont {Dirscherl}}, \bibinfo {author} {\bibfnamefont {A.}~\bibnamefont {Ianeselli}}, \bibinfo {author} {\bibfnamefont {A.}~\bibnamefont {K{\"u}hnlein}}, \bibinfo {author} {\bibfnamefont {K.}~\bibnamefont {Le~Vay}}, \bibinfo {author} {\bibfnamefont {P.}~\bibnamefont {Schwintek}}, \bibinfo {author} {\bibfnamefont {S.}~\bibnamefont {Islam}}, \bibinfo {author} {\bibfnamefont {M.~K.}\ \bibnamefont {Corpinot}}, \bibinfo {author} {\bibfnamefont {B.}~\bibnamefont {Scheu}}, \bibinfo {author} {\bibfnamefont {D.~B.}\ \bibnamefont {Dingwell}}, \bibinfo {author} {\bibfnamefont {P.}~\bibnamefont {Schwille}}, \bibinfo {author} {\bibfnamefont {H.}~\bibnamefont {Mutschler}}, \bibinfo {author} {\bibfnamefont {M.~W.}\ \bibnamefont {Powner}}, \bibinfo {author} {\bibfnamefont {C.~B.}\ \bibnamefont {Mast}},\ and\ \bibinfo {author}
  {\bibfnamefont {D.}~\bibnamefont {Braun}},\ }\bibfield  {title} {\bibinfo {title} {Heated gas bubbles enrich, crystallize, dry, phosphorylate and encapsulate prebiotic molecules},\ }\href {https://doi.org/10.1038/s41557-019-0299-5} {\bibfield  {journal} {\bibinfo  {journal} {Nature Chemistry}\ }\textbf {\bibinfo {volume} {11}},\ \bibinfo {pages} {779} (\bibinfo {year} {2019})}\BibitemShut {NoStop}%
\bibitem [{\citenamefont {Bartolucci}\ \emph {et~al.}(2023)\citenamefont {Bartolucci}, \citenamefont {Cala{\c c}a~Serr{\~a}o}, \citenamefont {Schwintek}, \citenamefont {K{\"u}hnlein}, \citenamefont {Rana}, \citenamefont {Janto}, \citenamefont {Hofer}, \citenamefont {Mast}, \citenamefont {Braun},\ and\ \citenamefont {Weber}}]{bartolucciSequenceSelfselectionCyclic2023}%
  \BibitemOpen
  \bibfield  {author} {\bibinfo {author} {\bibfnamefont {G.}~\bibnamefont {Bartolucci}}, \bibinfo {author} {\bibfnamefont {A.}~\bibnamefont {Cala{\c c}a~Serr{\~a}o}}, \bibinfo {author} {\bibfnamefont {P.}~\bibnamefont {Schwintek}}, \bibinfo {author} {\bibfnamefont {A.}~\bibnamefont {K{\"u}hnlein}}, \bibinfo {author} {\bibfnamefont {Y.}~\bibnamefont {Rana}}, \bibinfo {author} {\bibfnamefont {P.}~\bibnamefont {Janto}}, \bibinfo {author} {\bibfnamefont {D.}~\bibnamefont {Hofer}}, \bibinfo {author} {\bibfnamefont {C.~B.}\ \bibnamefont {Mast}}, \bibinfo {author} {\bibfnamefont {D.}~\bibnamefont {Braun}},\ and\ \bibinfo {author} {\bibfnamefont {C.~A.}\ \bibnamefont {Weber}},\ }\bibfield  {title} {\bibinfo {title} {Sequence self-selection by cyclic phase separation},\ }\href {https://doi.org/10.1073/pnas.2218876120} {\bibfield  {journal} {\bibinfo  {journal} {Proceedings of the National Academy of Sciences}\ }\textbf {\bibinfo {volume} {120}},\ \bibinfo {pages} {e2218876120} (\bibinfo {year} {2023})}\BibitemShut
  {NoStop}%
\bibitem [{\citenamefont {Unknown}()}]{LiberCoquina}%
  \BibitemOpen
  \bibfield  {author} {\bibinfo {author} {\bibnamefont {Unknown}},\ }\href@noop {} {\bibinfo {title} {Liber de coquina}},\ \bibinfo {note} {available at: \url{https://www.uni-giessen.de/de/fbz/fb05/germanistik/absprache/sprachverwendung/gloning/tx/mul2-lib.htm}}\BibitemShut {NoStop}%
\bibitem [{\citenamefont {Bertsch}\ \emph {et~al.}(2019)\citenamefont {Bertsch}, \citenamefont {Savorani},\ and\ \citenamefont {Fischer}}]{bertschRheologySwissCheese2019}%
  \BibitemOpen
  \bibfield  {author} {\bibinfo {author} {\bibfnamefont {P.}~\bibnamefont {Bertsch}}, \bibinfo {author} {\bibfnamefont {L.}~\bibnamefont {Savorani}},\ and\ \bibinfo {author} {\bibfnamefont {P.}~\bibnamefont {Fischer}},\ }\bibfield  {title} {\bibinfo {title} {Rheology of {{Swiss Cheese Fondue}}},\ }\href {https://doi.org/10.1021/acsomega.8b02424} {\bibfield  {journal} {\bibinfo  {journal} {ACS Omega}\ }\textbf {\bibinfo {volume} {4}},\ \bibinfo {pages} {1103} (\bibinfo {year} {2019})}\BibitemShut {NoStop}%
\bibitem [{\citenamefont {Bressanini}(2013)}]{bressaniniRicetteScientificheCacio}%
  \BibitemOpen
  \bibfield  {author} {\bibinfo {author} {\bibfnamefont {D.}~\bibnamefont {Bressanini}},\ }\href@noop {} {\bibinfo {title} {Le ricette scientifiche: La cacio e pepe}},\ \bibinfo {howpublished} {Le Scienze Blog} (\bibinfo {year} {2013}),\ \bibinfo {note} {available at: \url{ https://bressanini-lescienze.blogautore.espresso.repubblica.it/2013/05/13/le-ricette-scientifiche-la-cacio-e-pepe/}}\BibitemShut {NoStop}%
\bibitem [{\citenamefont {Anema}\ and\ \citenamefont {Li}(2003)}]{WheyMicelles1}%
  \BibitemOpen
  \bibfield  {author} {\bibinfo {author} {\bibfnamefont {S.~G.}\ \bibnamefont {Anema}}\ and\ \bibinfo {author} {\bibfnamefont {Y.}~\bibnamefont {Li}},\ }\bibfield  {title} {\bibinfo {title} {Association of denatured whey proteins with casein micelles in heated reconstituted skim milk and its effect on casein micelle size},\ }\href@noop {} {\bibfield  {journal} {\bibinfo  {journal} {Journal of dairy Research}\ }\textbf {\bibinfo {volume} {70}},\ \bibinfo {pages} {73} (\bibinfo {year} {2003})}\BibitemShut {NoStop}%
\bibitem [{\citenamefont {Pan}\ \emph {et~al.}(2022)\citenamefont {Pan}, \citenamefont {Ye}, \citenamefont {Dave}, \citenamefont {Fraser},\ and\ \citenamefont {Singh}}]{WheyMicelles2}%
  \BibitemOpen
  \bibfield  {author} {\bibinfo {author} {\bibfnamefont {Z.}~\bibnamefont {Pan}}, \bibinfo {author} {\bibfnamefont {A.}~\bibnamefont {Ye}}, \bibinfo {author} {\bibfnamefont {A.}~\bibnamefont {Dave}}, \bibinfo {author} {\bibfnamefont {K.}~\bibnamefont {Fraser}},\ and\ \bibinfo {author} {\bibfnamefont {H.}~\bibnamefont {Singh}},\ }\bibfield  {title} {\bibinfo {title} {Kinetics of heat-induced interactions among whey proteins and casein micelles in sheep skim milk and aggregation of the casein micelles},\ }\href@noop {} {\bibfield  {journal} {\bibinfo  {journal} {Journal of Dairy Science}\ }\textbf {\bibinfo {volume} {105}},\ \bibinfo {pages} {3871} (\bibinfo {year} {2022})}\BibitemShut {NoStop}%
\bibitem [{\citenamefont {Bauermann}\ \emph {et~al.}(2024)\citenamefont {Bauermann}, \citenamefont {Bartolucci}, \citenamefont {Boekhoven}, \citenamefont {J{\"u}licher},\ and\ \citenamefont {Weber}}]{bauermannCriticalTransitionIntensive2024}%
  \BibitemOpen
  \bibfield  {author} {\bibinfo {author} {\bibfnamefont {J.}~\bibnamefont {Bauermann}}, \bibinfo {author} {\bibfnamefont {G.}~\bibnamefont {Bartolucci}}, \bibinfo {author} {\bibfnamefont {J.}~\bibnamefont {Boekhoven}}, \bibinfo {author} {\bibfnamefont {F.}~\bibnamefont {J{\"u}licher}},\ and\ \bibinfo {author} {\bibfnamefont {C.~A.}\ \bibnamefont {Weber}},\ }\href {https://doi.org/10.48550/arXiv.2409.03629} {\bibinfo {title} {Critical transition between intensive and extensive active droplets}} (\bibinfo {year} {2024}),\ \Eprint {https://arxiv.org/abs/2409.03629} {arXiv:2409.03629} \BibitemShut {NoStop}%
\bibitem [{\citenamefont {Bray}(1994)}]{brayTheoryPhaseorderingKinetics1994}%
  \BibitemOpen
  \bibfield  {author} {\bibinfo {author} {\bibfnamefont {A.}~\bibnamefont {Bray}},\ }\bibfield  {title} {\bibinfo {title} {Theory of phase-ordering kinetics},\ }\href {https://doi.org/10.1080/00018739400101505} {\bibfield  {journal} {\bibinfo  {journal} {Advances in Physics}\ }\textbf {\bibinfo {volume} {43}},\ \bibinfo {pages} {357} (\bibinfo {year} {1994})}\BibitemShut {NoStop}%
\bibitem [{\citenamefont {Donovan}\ and\ \citenamefont {Mulvihill}(1987)}]{donovan1987thermal}%
  \BibitemOpen
  \bibfield  {author} {\bibinfo {author} {\bibfnamefont {M.}~\bibnamefont {Donovan}}\ and\ \bibinfo {author} {\bibfnamefont {D.}~\bibnamefont {Mulvihill}},\ }\bibfield  {title} {\bibinfo {title} {Thermal denaturation and aggregation of whey proteins},\ }\href@noop {} {\bibfield  {journal} {\bibinfo  {journal} {Irish Journal of Food Science and Technology}\ ,\ \bibinfo {pages} {87}} (\bibinfo {year} {1987})}\BibitemShut {NoStop}%
\bibitem [{\citenamefont {Zhang}\ \emph {et~al.}(2021)\citenamefont {Zhang}, \citenamefont {Zhou}, \citenamefont {Zhang},\ and\ \citenamefont {Zhou}}]{zhang2021heat}%
  \BibitemOpen
  \bibfield  {author} {\bibinfo {author} {\bibfnamefont {L.}~\bibnamefont {Zhang}}, \bibinfo {author} {\bibfnamefont {R.}~\bibnamefont {Zhou}}, \bibinfo {author} {\bibfnamefont {J.}~\bibnamefont {Zhang}},\ and\ \bibinfo {author} {\bibfnamefont {P.}~\bibnamefont {Zhou}},\ }\bibfield  {title} {\bibinfo {title} {Heat-induced denaturation and bioactivity changes of whey proteins},\ }\href@noop {} {\bibfield  {journal} {\bibinfo  {journal} {International Dairy Journal}\ }\textbf {\bibinfo {volume} {123}},\ \bibinfo {pages} {105175} (\bibinfo {year} {2021})}\BibitemShut {NoStop}%
\bibitem [{\citenamefont {Flory}(1942)}]{floryThermodynamicsHighPolymer1942}%
  \BibitemOpen
  \bibfield  {author} {\bibinfo {author} {\bibfnamefont {P.~J.}\ \bibnamefont {Flory}},\ }\bibfield  {title} {\bibinfo {title} {Thermodynamics of {{High Polymer Solutions}}},\ }\href@noop {} {\bibfield  {journal} {\bibinfo  {journal} {The Journal of Chemical Physics}\ }\textbf {\bibinfo {volume} {10}},\ \bibinfo {pages} {51} (\bibinfo {year} {1942})}\BibitemShut {NoStop}%
\bibitem [{\citenamefont {Huggins}(1942)}]{hugginsThermodynamicPropertiesSolutions1942}%
  \BibitemOpen
  \bibfield  {author} {\bibinfo {author} {\bibfnamefont {M.~L.}\ \bibnamefont {Huggins}},\ }\bibfield  {title} {\bibinfo {title} {Thermodynamic properties of solutions of long-chain compounds},\ }\href@noop {} {\bibfield  {journal} {\bibinfo  {journal} {Annals of the New York Academy of Sciences}\ }\textbf {\bibinfo {volume} {43}},\ \bibinfo {pages} {1} (\bibinfo {year} {1942})}\BibitemShut {NoStop}%
\bibitem [{\citenamefont {Safran}(2019)}]{safranStatisticalThermodynamicsSurfaces2019}%
  \BibitemOpen
  \bibfield  {author} {\bibinfo {author} {\bibfnamefont {S.~A.}\ \bibnamefont {Safran}},\ }\bibfield  {title} {\bibinfo {title} {Statistical thermodynamics of surfaces, interfaces, and membranes},\ }\bibfield  {journal} {\bibinfo  {journal} {CRC Press}\ }\href {https://doi.org/10.1201/9780429497131} {10.1201/9780429497131} (\bibinfo {year} {2019})\BibitemShut {NoStop}%
\bibitem [{\citenamefont {Fritsch}\ \emph {et~al.}(2021)\citenamefont {Fritsch}, \citenamefont {{Diaz-Delgadillo}}, \citenamefont {{Adame-Arana}}, \citenamefont {Hoege}, \citenamefont {Mittasch}, \citenamefont {Kreysing}, \citenamefont {Leaver}, \citenamefont {Hyman}, \citenamefont {J{\"u}licher},\ and\ \citenamefont {Weber}}]{fritschLocalThermodynamicsGovern2021}%
  \BibitemOpen
  \bibfield  {author} {\bibinfo {author} {\bibfnamefont {A.~W.}\ \bibnamefont {Fritsch}}, \bibinfo {author} {\bibfnamefont {A.~F.}\ \bibnamefont {{Diaz-Delgadillo}}}, \bibinfo {author} {\bibfnamefont {O.}~\bibnamefont {{Adame-Arana}}}, \bibinfo {author} {\bibfnamefont {C.}~\bibnamefont {Hoege}}, \bibinfo {author} {\bibfnamefont {M.}~\bibnamefont {Mittasch}}, \bibinfo {author} {\bibfnamefont {M.}~\bibnamefont {Kreysing}}, \bibinfo {author} {\bibfnamefont {M.}~\bibnamefont {Leaver}}, \bibinfo {author} {\bibfnamefont {A.~A.}\ \bibnamefont {Hyman}}, \bibinfo {author} {\bibfnamefont {F.}~\bibnamefont {J{\"u}licher}},\ and\ \bibinfo {author} {\bibfnamefont {C.~A.}\ \bibnamefont {Weber}},\ }\bibfield  {title} {\bibinfo {title} {Local thermodynamics govern formation and dissolution of {{Caenorhabditis}} elegans {{P}} granule condensates},\ }\href {https://doi.org/10.1073/pnas.2102772118} {\bibfield  {journal} {\bibinfo  {journal} {Proceedings of the National Academy of Sciences}\ }\textbf {\bibinfo {volume} {118}},\
  \bibinfo {pages} {e2102772118} (\bibinfo {year} {2021})}\BibitemShut {NoStop}%
\bibitem [{\citenamefont {Anema}(2021)}]{anema2021heat}%
  \BibitemOpen
  \bibfield  {author} {\bibinfo {author} {\bibfnamefont {S.~G.}\ \bibnamefont {Anema}},\ }\bibfield  {title} {\bibinfo {title} {Heat-induced changes in caseins and casein micelles, including interactions with denatured whey proteins},\ }\href@noop {} {\bibfield  {journal} {\bibinfo  {journal} {International Dairy Journal}\ }\textbf {\bibinfo {volume} {122}},\ \bibinfo {pages} {105136} (\bibinfo {year} {2021})}\BibitemShut {NoStop}%
\bibitem [{\citenamefont {Nicolai}\ and\ \citenamefont {Chassenieux}(2021)}]{Micelle1}%
  \BibitemOpen
  \bibfield  {author} {\bibinfo {author} {\bibfnamefont {T.}~\bibnamefont {Nicolai}}\ and\ \bibinfo {author} {\bibfnamefont {C.}~\bibnamefont {Chassenieux}},\ }\bibfield  {title} {\bibinfo {title} {Heat-induced gelation of casein micelles},\ }\href@noop {} {\bibfield  {journal} {\bibinfo  {journal} {Food Hydrocolloids}\ }\textbf {\bibinfo {volume} {118}},\ \bibinfo {pages} {106755} (\bibinfo {year} {2021})}\BibitemShut {NoStop}%
\bibitem [{\citenamefont {Liu}\ \emph {et~al.}(2023)\citenamefont {Liu}, \citenamefont {Tan}, \citenamefont {Xu}, \citenamefont {Liu}, \citenamefont {Zhu},\ and\ \citenamefont {Dong}}]{Complex1}%
  \BibitemOpen
  \bibfield  {author} {\bibinfo {author} {\bibfnamefont {Y.}~\bibnamefont {Liu}}, \bibinfo {author} {\bibfnamefont {Z.}~\bibnamefont {Tan}}, \bibinfo {author} {\bibfnamefont {X.}~\bibnamefont {Xu}}, \bibinfo {author} {\bibfnamefont {J.}~\bibnamefont {Liu}}, \bibinfo {author} {\bibfnamefont {B.}~\bibnamefont {Zhu}},\ and\ \bibinfo {author} {\bibfnamefont {X.}~\bibnamefont {Dong}},\ }\bibfield  {title} {\bibinfo {title} {The formation of rice starch/casein complex by hydrothermal and stabilizing high internal phase emulsions with 3d printing property},\ }\href@noop {} {\bibfield  {journal} {\bibinfo  {journal} {Food Hydrocolloids}\ }\textbf {\bibinfo {volume} {144}},\ \bibinfo {pages} {108995} (\bibinfo {year} {2023})}\BibitemShut {NoStop}%
\bibitem [{\citenamefont {Monosilio}(2023)}]{YT_video}%
  \BibitemOpen
  \bibfield  {author} {\bibinfo {author} {\bibfnamefont {L.}~\bibnamefont {Monosilio}},\ }\href@noop {} {\bibinfo {title} {Cacio e pepe: originale vs. infallibile vs. gourmet con luciano monosilio | babish action review}},\ \bibinfo {howpublished} {YouTube} (\bibinfo {year} {2023}),\ \bibinfo {note} {available at: \url{https://www.youtube.com/watch?v=U4eaNqTbDDA}}\BibitemShut {NoStop}%
\bibitem [{\citenamefont {Bartolucci}\ \emph {et~al.}(2021)\citenamefont {Bartolucci}, \citenamefont {{Adame-Arana}}, \citenamefont {Zhao},\ and\ \citenamefont {Weber}}]{bartolucciControllingCompositionCoexisting2021}%
  \BibitemOpen
  \bibfield  {author} {\bibinfo {author} {\bibfnamefont {G.}~\bibnamefont {Bartolucci}}, \bibinfo {author} {\bibfnamefont {O.}~\bibnamefont {{Adame-Arana}}}, \bibinfo {author} {\bibfnamefont {X.}~\bibnamefont {Zhao}},\ and\ \bibinfo {author} {\bibfnamefont {C.~A.}\ \bibnamefont {Weber}},\ }\bibfield  {title} {\bibinfo {title} {Controlling composition of coexisting phases via molecular transitions},\ }\href {https://doi.org/10.1016/j.bpj.2021.09.036} {\bibfield  {journal} {\bibinfo  {journal} {Biophysical Journal}\ }\textbf {\bibinfo {volume} {120}},\ \bibinfo {pages} {4682} (\bibinfo {year} {2021})}\BibitemShut {NoStop}%
\bibitem [{\citenamefont {Israelachvili}(2015)}]{israelachviliIntermolecularSurfaceForces2015}%
  \BibitemOpen
  \bibfield  {author} {\bibinfo {author} {\bibfnamefont {J.~N.}\ \bibnamefont {Israelachvili}},\ }\href@noop {} {\emph {\bibinfo {title} {Intermolecular and {{Surface Forces}}}}}\ (\bibinfo  {publisher} {Academic Press},\ \bibinfo {year} {2015})\BibitemShut {NoStop}%
\bibitem [{\citenamefont {Deviri}\ and\ \citenamefont {Safran}(2020)}]{deviriEquilibriumSizeDistribution2020}%
  \BibitemOpen
  \bibfield  {author} {\bibinfo {author} {\bibfnamefont {D.}~\bibnamefont {Deviri}}\ and\ \bibinfo {author} {\bibfnamefont {S.~A.}\ \bibnamefont {Safran}},\ }\bibfield  {title} {\bibinfo {title} {Equilibrium size distribution and phase separation of multivalent, molecular assemblies in dilute solution},\ }\href {https://doi.org/10.1039/C9SM02408E} {\bibfield  {journal} {\bibinfo  {journal} {Soft Matter}\ }\textbf {\bibinfo {volume} {16}},\ \bibinfo {pages} {5458} (\bibinfo {year} {2020})}\BibitemShut {NoStop}%
\bibitem [{\citenamefont {Bartolucci}\ \emph {et~al.}(2024)\citenamefont {Bartolucci}, \citenamefont {Haugerud}, \citenamefont {Michaels},\ and\ \citenamefont {Weber}}]{Bartolucci2024}%
  \BibitemOpen
  \bibfield  {author} {\bibinfo {author} {\bibfnamefont {G.}~\bibnamefont {Bartolucci}}, \bibinfo {author} {\bibfnamefont {I.~S.}\ \bibnamefont {Haugerud}}, \bibinfo {author} {\bibfnamefont {T.~C.}\ \bibnamefont {Michaels}},\ and\ \bibinfo {author} {\bibfnamefont {C.~A.}\ \bibnamefont {Weber}},\ }\bibfield  {title} {\bibinfo {title} {The interplay between biomolecular assembly and phase separation},\ }\bibfield  {journal} {\bibinfo  {journal} {eLife}\ }\href {https://doi.org/10.7554/elife.93003.2} {10.7554/elife.93003.2} (\bibinfo {year} {2024})\BibitemShut {NoStop}%
\end{thebibliography}%



\setcounter{equation}{0}
\setcounter{figure}{0}
\setcounter{table}{0}
\setcounter{page}{0}
\setcounter{section}{0}
\makeatletter
\numberwithin{equation}{section}
\renewcommand{\thesection}{S\arabic{section}}
\renewcommand{\theequation}{S\arabic{equation}}
\renewcommand{\thetable}{S\arabic{table}}
\renewcommand{\thefigure}{S\arabic{figure}}


\onecolumngrid

\begin{center}

\newpage
{\LARGE Supplementary Material for}\\
\vspace{0.4cm}
\textbf{\Large Phase behavior of Cacio e Pepe sauce}\\
\end{center}

\section{Equipment, experimental protocol and ingredients}\label{sec:S1}
\subsection*{Equipment}
The equipment utilized in this study can be categorized into two functional components: the sauce preparation and heating system (Figures S1a and S2a), and the image acquisition system (Figures S1b and S2b).

\hspace{0.5cm}
\begin{figure*}[!h] 
\includegraphics[width=0.95\linewidth]{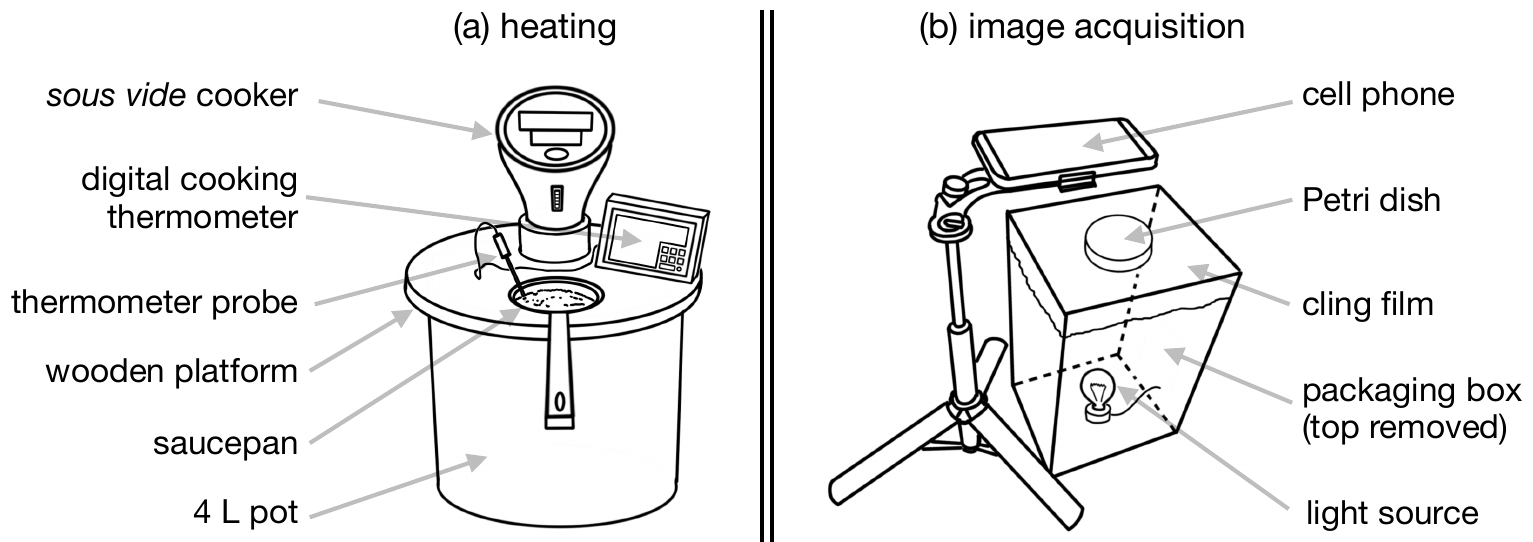}
\caption{Scheme of the experimental equipment used in this work:
(a) heating equipment, and
(b) image acquisition equipment.}
\label{fig:schema_apparato}
\end{figure*} 

\normalem

The sauce was prepared using standard kitchen tools, including a kitchen scale, mixing cups, an immersion blender, and spoons. 
The sauce heating process utilized a \emph{sous vide} cooker in a `modified setting', which included a custom-built wooden platform designed to support the cooker in the correct position and facilitate the immersion of the saucepan containing the sauce into temperature-controlled water. 
The hole in the wooden platform was specifically designed to serve as a constraint for the saucepan, preventing it from floating on the water of the pot, which would otherwise reduce the efficiency of heat transfer.


The image acquisition setup employed a cell phone equipped with a 12 MP f/1.6 camera (iPhone 13). 
The phone was mounted on a tripod positioned above a custom-made transparent support designed to hold the Petri dish containing the sauce sample. 
This support was constructed by removing the top portion of a cardboard packaging box and replacing it with transparent cling film, onto which the sample was deposited. 
A table lamp inserted through an opening at the bottom of the box served as a light source, illuminating the sample from below so that the clumps of condensed sauce appear as darker spots in the resulting images.

\ULforem

\subsection*{Experimental protocol}
The experimental protocol we followed in this work can be summarized in the following 
steps:
\begin{enumerate}
\item \textbf{Sauce preparation}: 
the ingredients for a small batch of sauce ($\approx 110$g, with the exact weight depending on the specific formulation) are mixed in a clean mug to achieve the desired composition of cacio cheese, starch, and water. 
The mixture is homogenized using an immersion blender, then transferred into a saucepan and weighed. 
The saucepan is then placed into the wooden platform of the heating apparatus (\fig{schema_apparato}(a)).

\item \textbf{Temperature ramp}: 
the sauce is gradually heated up
while its actual temperature is continuously monitored using a digital cooking thermometer with a probe immersed
in the sauce itself. 
During the heating process, the mixture is constantly stirred  with a spoon to ensure uniform heating and to prevent the formation of large cheese aggregates on the saucepan walls.


When the target temperature is reached, the saucepan is promptly removed and weighed. If a detectable weight loss due to evaporation is observed (typically 1–2g), the lost amount of water is replenished by adding the exact quantity withdrawn from the water in the pot, which is approximately at the same temperature as the sauce. The saucepan is then returned to the platform, stirred thoroughly, and the temperature is rechecked. A slight decrease in temperature may occur due to heat dissipation during the weighing process. Once the temperature returns to the desired value (usually within a few seconds under stirring), a sample is taken from the saucepan for image acquisition. The saucepan is then weighed again to establish the reference point for the next experimental data point in the temperature ramp.

\item \textbf{Image acquisition}: 
the sauce sample withdrawn from the saucepan is quickly transferred into a clean Petri dish. 
The mixture is evenly spread across the dish by gently shaking it, and then the dish is positioned on the transparent support (\fig{schema_apparato}(b)). 
A photograph of the sample is captured using the cell phone. 
After imaging, the sauce sample is collected in a separate container for later consumption. 
This entire operation is optimized to be completed within approximately 20 seconds, ensuring that no significant precipitation of cheese clumps occurs during 
the sample's transfer and photography.




\item \textbf{Next temperature}: steps 2 and 3 are repeated for all target temperatures to construct an experimental phase diagram. 
The amount of sauce prepared for each batch is planned to be 
enough for all the measurements, with minimal leftovers at the end of the ramp (any extra is consumed afterward).



\end{enumerate}
To ensure the statistical soundness of our conclusions, each temperature ramp has been replicated a minimum of twice with exactly the same control parameters.

\begin{figure*}[!h] 
\includegraphics[width=0.75\linewidth]{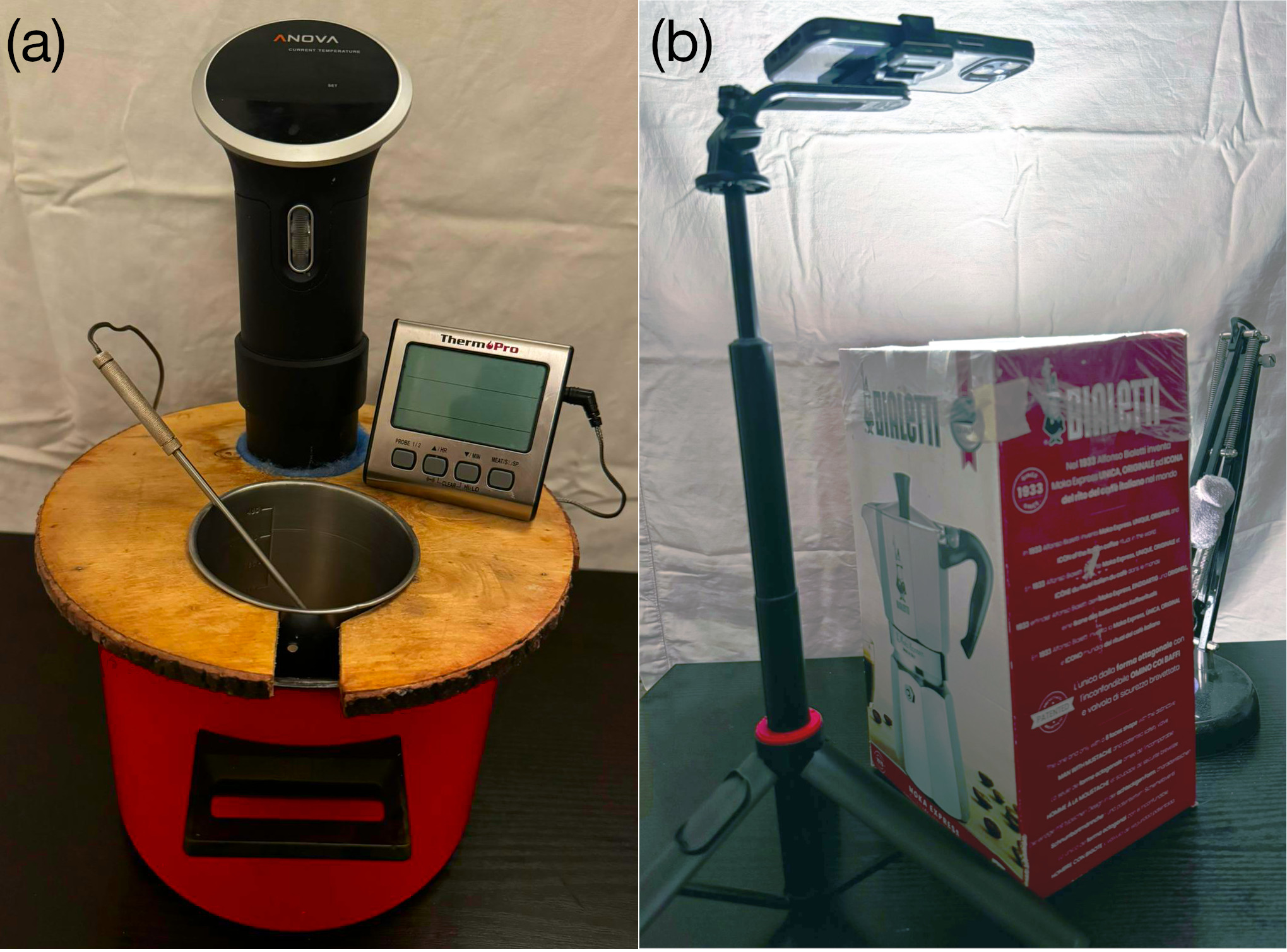}
\caption{Photos of the actual experimental equipment used in this work:
(a) heating part and 
(b) image acquisition part.
}
\label{fig:foto_apparato}
\end{figure*}

\subsection*{Ingredients} \label{Ingredients}
\begin{figure}
    \centering
\includegraphics[width=.25\linewidth]{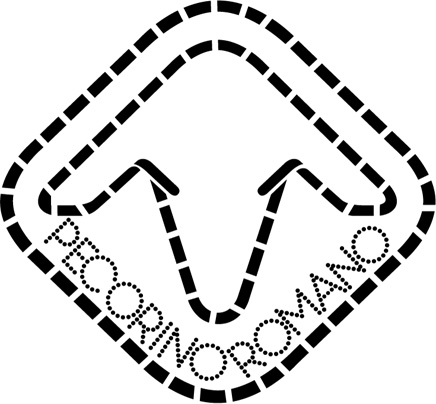}
\includegraphics[width=.25\linewidth]{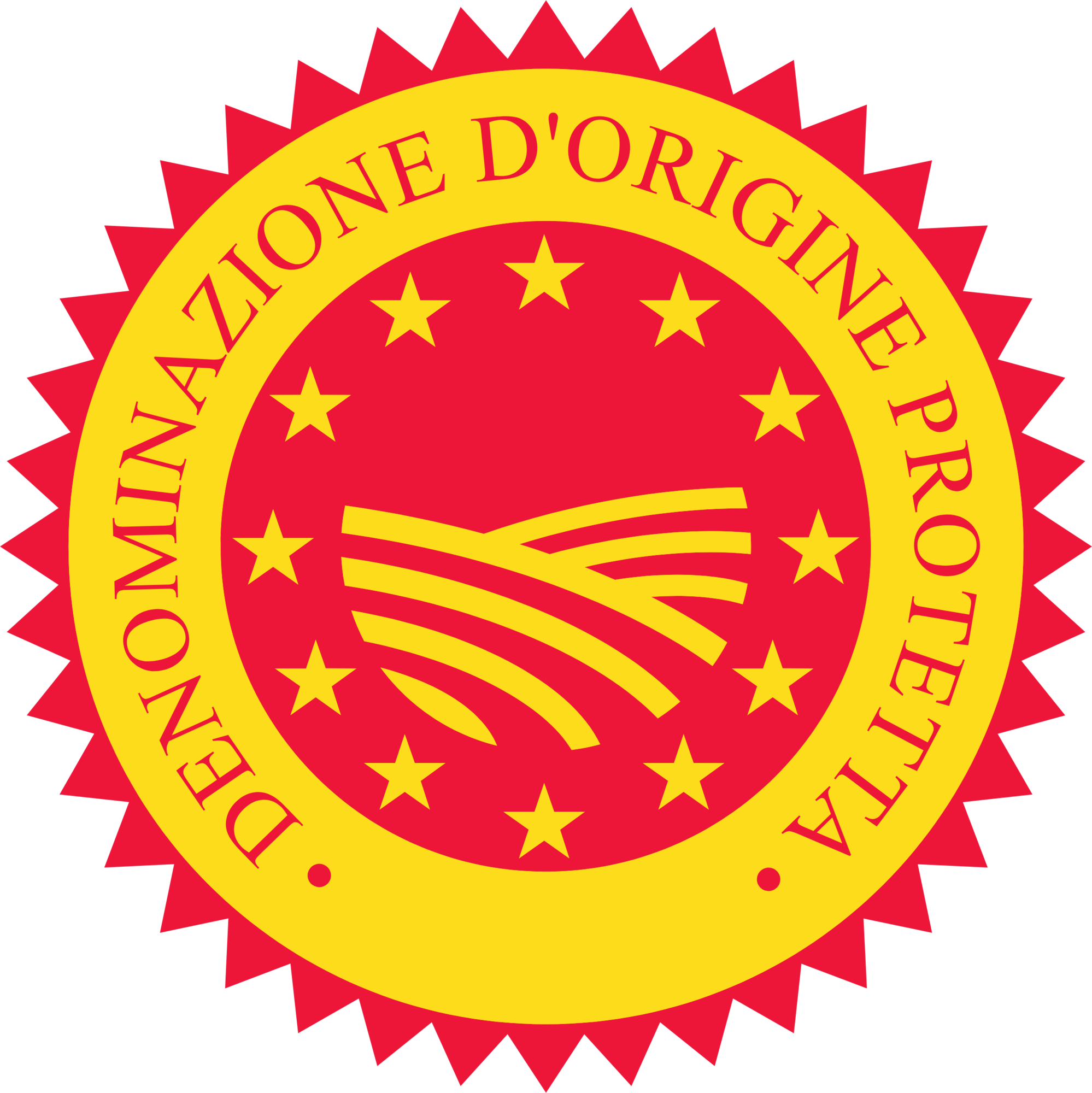} 
        \caption{Official logo of Pecorino Romano DOP and original DOP products}
       \label{fig:logo}
\end{figure}

\tcr{The two ingredients 
employed in this work are dry corn starch, which was purchased in a common supermarket, and the pecorino cheese.
For the latter, we opted for a Pecorino Romano DOP which has not to be confused with a common romano cheese and can be recognized by the logo in \fig{logo}. The acronym DOP, which stands for "Denominazione di Origine Protetta" (protected designation of origin), is a European Union designation which indicates products that have been produced, processed and developed in a specific geographical area, using the recognized know-how of local producers and ingredients from the region concerned.
This fact guarantees the quality of the cheese and its abiding to strict rules of production, in particular a defined list of ingredients.
For our work we selected a Pecorino Romano DOP from Sardinia, Italy. Indeed, even if the name of the cheese indicates Lazio as the origin of the product, nowadays most of its production is done in the island of Sardinia.
The complete nutritional information of the cheese used in our work is reported in Table \ref{tab:nutritional}.}

\vspace{1cm}
\begin{table}[h]
\begin{tabularx}{0.35\textwidth}
{>{\centering\arraybackslash}X|>{\centering\arraybackslash}X}
\hline
\hline
Energy & 1643 kJ / 396 kcal\\
\hline
\hline
Fat & 32 g \\
\hline
of which saturated & 20 g \\
\hline
\hline
Carbohydrates &  < 0.50 g\\
\hline
of which sugars &  < 0.50 g\\
\hline
\hline
Protein & 27 g \\
\hline
\hline
Salt &  3.5 g\\
\hline
\hline
\end{tabularx}
\caption{Nutritional declaration of the cacio cheese employed in this work.}
\label{tab:nutritional}
\end{table}

\section{Data analysis}\label{sec:data_analysis}

The data analysis involved segmenting aggregates in petri dish images and computing their \tcr{size}. Initially, the images were converted to greyscale and normalized to an intensity range of \([0, 1]\). To enhance visibility, we adjusted exposure using the \texttt{equalize\_adapthist} function from the \texttt{exposure} module in \texttt{scikit-image}, with a clip limit of \(0.01\). Images were then smoothed with a Gaussian filter using a sigma of \(1\) pixel. These parameters were determined based on iterative trials.

Next, images were cropped around the center of the petri dish to ensure uniformity in size across samples. A quantile-based segmentation was then applied. Specifically, a quantile value was selected for each sample within the interval \([0.05, 0.25]\). The image was binarized as 
\[
M_{ij} = I_{ij} < I_q,
\]
where \(I_q\) is the intensity at the selected quantile $q$. The binary mask \(M_{ij}\) served as the seed for the watershed algorithm from \texttt{scikit-image}. For samples where aggregates formed a single large blob, the inverted image, \(1 - I_{ij}\), was used instead.

\begin{figure}
    \centering
\includegraphics[width=0.5\linewidth]{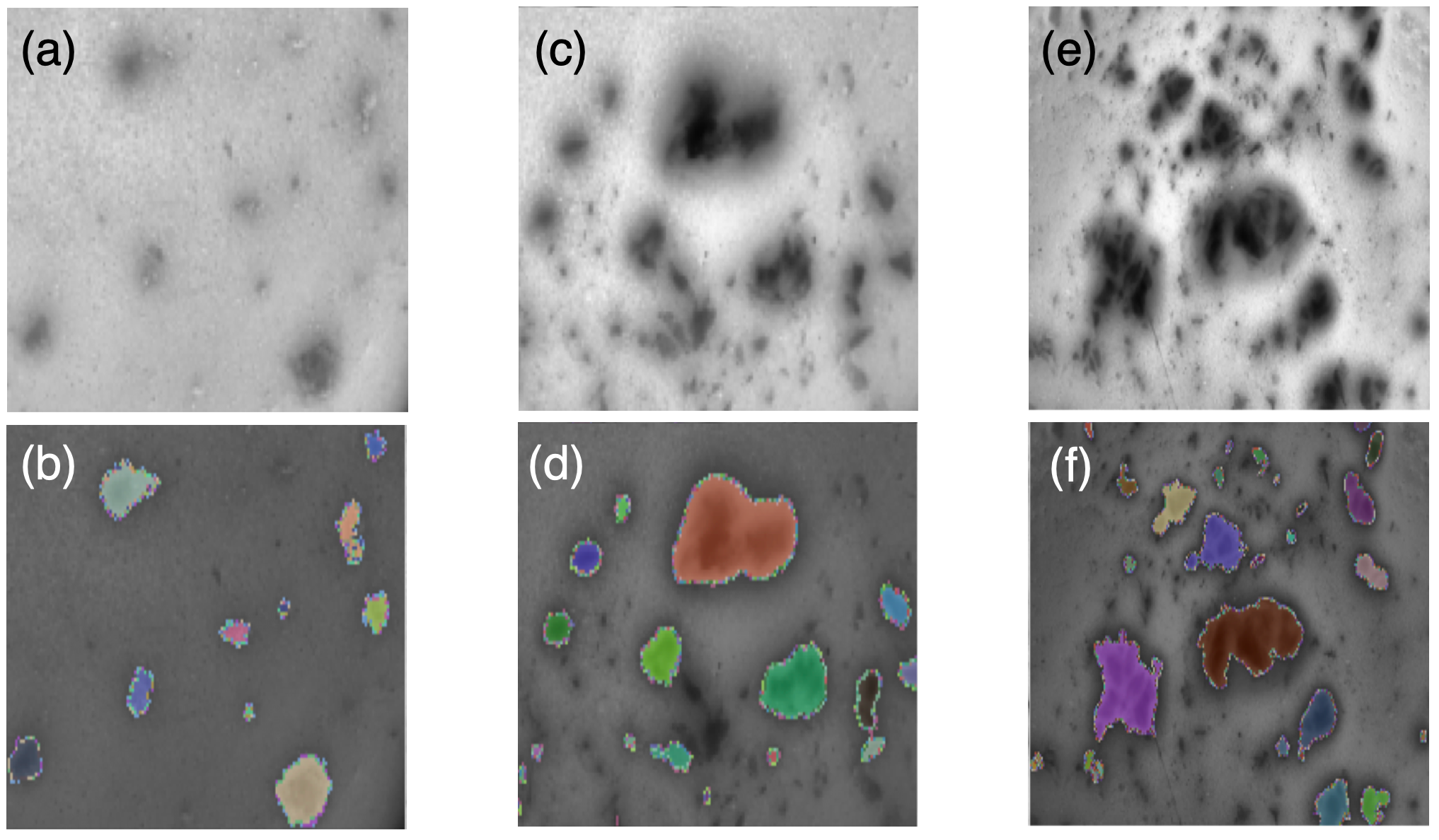} 
        \caption{Three examples of segmented images illustrating varying degrees of separation: (a), (b) low, (c), (d) medium, and (e), (f) high. Panels (a), (c), and (e) show cropped grayscale experimental images (not to scale), where darker regions represent aggregates. Corresponding segmented aggregates are displayed in panels (b), (d), and (f), obtained using the histogram-based segmentation method described below.}
       \label{fig:suppcrop}
\end{figure}

Starting from the watershed results, we labeled the segmented regions and applied standard post-processing steps, including small object removal to address artifacts, binary dilation, hole filling, and removal of regions touching the image border.

The segmented aggregates were then analyzed using the \texttt{regionprops} routine from \texttt{scikit-image}. In particular, we reported the mean \tcr{aggregate size} in the phase diagrams, calculated using the major axis length property from \texttt{regionprops}.

To \tcr{compute the major axis length, which we use as a proxy for the aggregate size}, one first calculates the covariance matrix \(C_{ab}\) for the pixel coordinates \(\{x_i, y_i\}\) of each aggregate, using the following flat measure:
\[
C_{xy} = \frac{1}{N} \sum_i x_i y_i - \frac{1}{N^2}  \sum_{ij} x_i y_j ,
\]
with analogous formulas for \(C_{xx}\) and \(C_{yy}\). A covariance matrix carries the information of the aggregate area as $\sqrt{\det C}$. However, this area differs from the actual aggregate area, and so the matrix \(C_{ab}\) was scaled such that \(\det C = A^2\), where \(A\) is the area of the segmented region. The largest eigenvalue of the rescaled \(C_{ab}\) corresponds to the square of the aggregate's \tcr{size}.

\begin{figure}[H]
    \centering
    \includegraphics[width=0.5\linewidth]{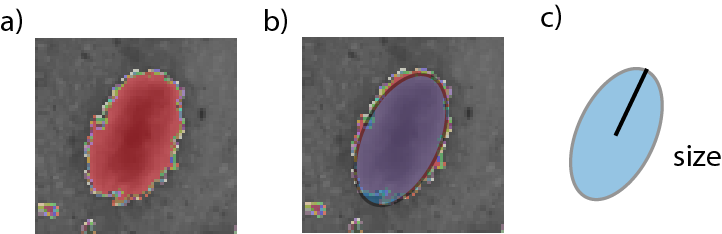}
    \caption{Visual guide to understand how an ellipse is used to measure an aggregate size. a) Segmented aggregate. b) Best ellipse overlaying the aggregate. c) Visual guide to identify the ellipse size via its major axis length.}
    \label{fig:ellipse}
\end{figure}

\begin{figure}[H]
    \centering
    \includegraphics[width=1\linewidth]{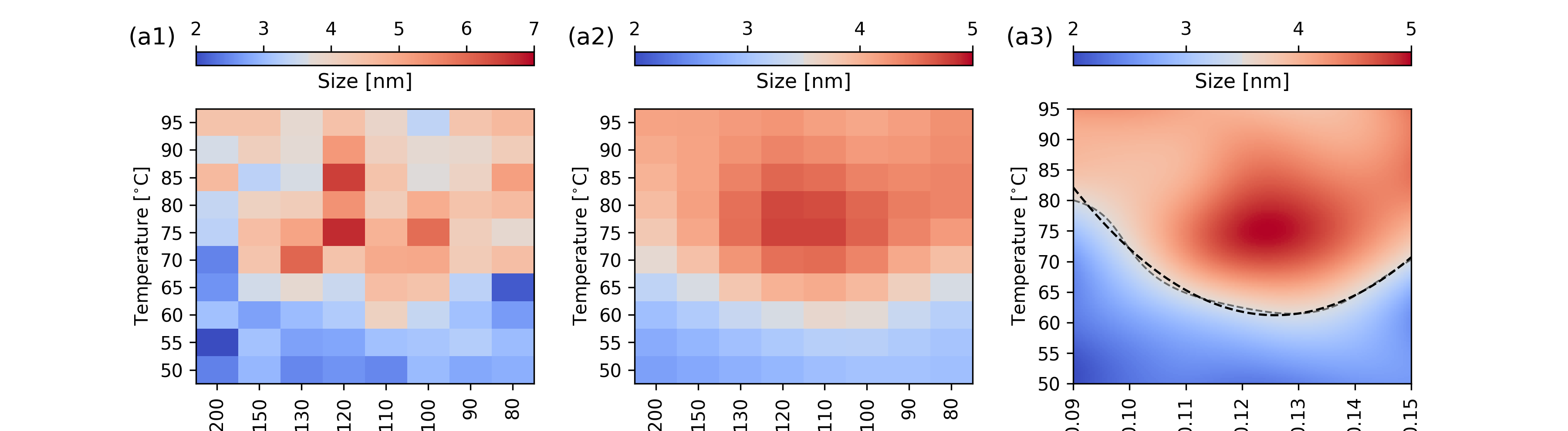}
    \vspace{0.2cm}\\
    \includegraphics[width=1\linewidth]{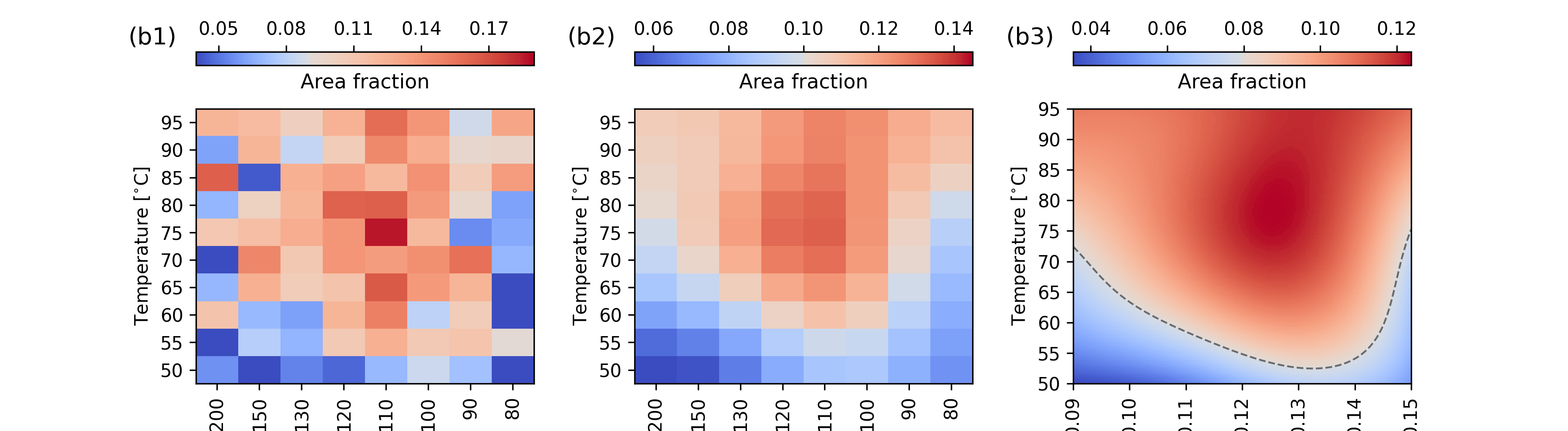}
    \vspace{0.2cm}\\
    \includegraphics[width=1\linewidth]{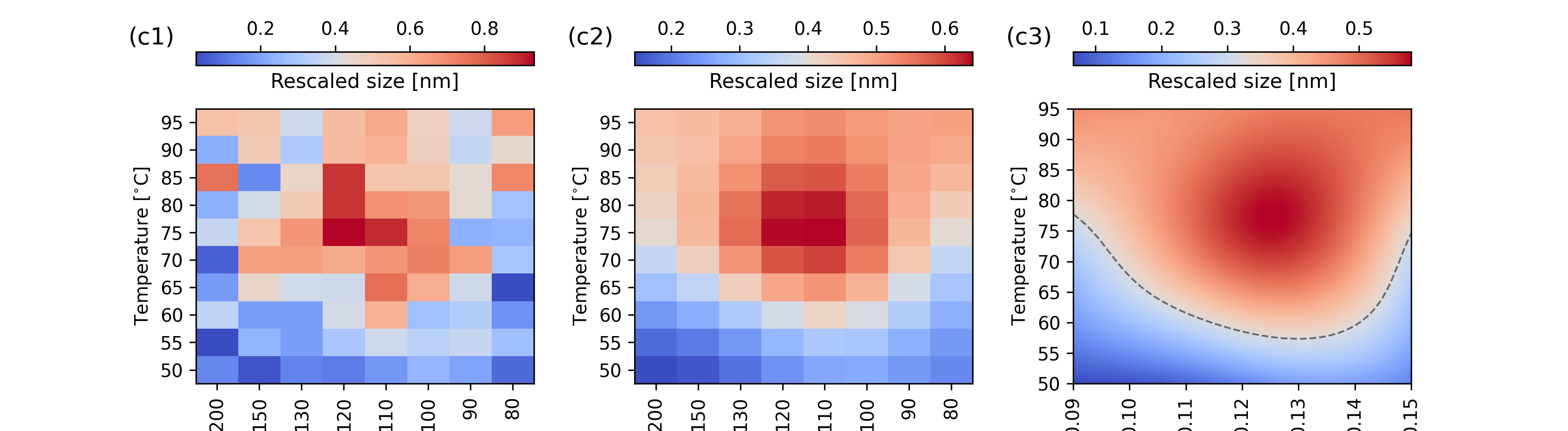}
    \caption{\tcr{Quantification of the degree of separation of the data in Fig.~\ref{fig:vary_water} using different order parameters. In (a1)-(a3) the major axis of the ellipses averaged over the clumps is displayed, like in the main text. We compare it with the ellipse's area (b1)-(b3) and the rescaled size, i.e. the product of area fraction times the average values of the ellipses's major axes (c1)-(c3). For the different order parameters, we show the raw data (a1)-(c1), data after Gaussian filtering (a2)-(c2), and kernel regression smoothing (a3)-(c3).}}
    \label{fig:compare_order_p}
\end{figure}

\tcr{In the main text, we used the major axis of the detected ellipses as a proxy for clump sizes. The average of this quantity over all clumps served as our order parameter, allowing us to assess the degree of separation in the mixture. For completeness, we compare phase diagrams constructed using different order parameters. In particular, in addition to the ellipses' major axis (\fig{compare_order_p}(a)), we also consider the area fraction (\fig{compare_order_p}(b)) and a ``rescaled size'' (\fig{compare_order_p}(c)). The area fraction is defined as the total area occupied by aggregates divided by the total sample area, properly restricted to the Petri dish region. The rescaled size, on the other hand, is obtained by multiplying the average major axis length of the ellipses by the area fraction. This measure was introduced to mitigate the impact of a single large clump dominating the statistics. Importantly, notice that all size measures lead to qualitatively similar phase diagrams.}

\section{Some analytical formulas}

Here we explicitly show Eq. \eqref{chi_and_n} as a function of $\phi^{\rm I}$ and $\phi^{\rm II}$, namely: 
\begin{align}\label{chi_and_n_explicit}
\chi(T) = & ~\mathcal{F}(\phi^\text{I},\phi^\text{II})= \frac{ \left(1 + \ln\left(\phi^{\rm I}\right)\right) \ln\left(1 - \phi^{\rm II}\right) - \left(1 + \ln\left(1 - \phi^{\rm I}\right)\right) \ln\left(\phi^{\rm II}\right)-2 \operatorname{Tanh}^{\rm -1}\left(1 - 2 \phi^{\rm I}\right)}{2 (\phi^{\rm I} -  \phi^{\rm II}) + (1 - 2 \phi^{\rm II} )\ln\left(\phi^{\rm I}\right) - \left(1 - 2 \phi^{\rm I}\right) \ln\left(\phi^{\rm II}\right)} \\[10pt]
n(T) = & ~\mathcal{G}(\phi^\text{I},\phi^\text{II}) =  \frac{2 (\phi^{\rm I} - \phi^{\rm II}) + (1 - 2 \phi^{\rm II})\ln\left(\phi^{\rm I}\right) - \left(1 - 2 \phi^{\rm I}\right) \ln\left(\phi^{\rm II}\right)}{2 \left(\phi^{\rm I} - \phi^{\rm II}\right) + \left(1 - 2 \phi^{\rm II}\right) \ln\left(1 - \phi^{\rm I}\right) - \left(1 - 2 \phi^{\rm I}\right) \ln\left(1 - \phi^{\rm II}\right)}
\end{align}

\end{document}